\documentclass[longauth]{aa} 
\makeatletter
\let\linenumbers\relax

\makeatother

\usepackage{txfonts}
\usepackage{graphicx,epsf}
\usepackage{amsmath,graphicx,adjustbox,setspace, sidecap, float}
\usepackage[toc,page]{appendix}
\usepackage{gensymb}
\usepackage{hyperref}
\usepackage{fancyhdr}
\usepackage{natbib}
\usepackage{caption,subcaption}
\usepackage[table, dvipsnames]{xcolor}
\usepackage{physics}
\usepackage{ulem}
\usepackage{multirow}
\usepackage{orcidlink}
\usepackage{placeins}

\newcommand{\sect}[1]{Sect.~\ref{sec:#1}}
\newcommand{\Sect}[1]{Section~\ref{sec:#1}}

\newcommand{\fg}[1]{Fig.~\ref{fig:#1}}
\newcommand{\Fg}[1]{Figure~\ref{fig:#1}}

\newcommand{\Fgs}[2]{Figures\ \ref{fig:#1} and \ref{fig:#2}}

\begin{document}
\title{MINDS: The very low-mass star and brown dwarf sample}

\subtitle{II. Probing disk settling, dust properties, and dust-gas interplay with JWST/MIRI}

   \author{Hyerin Jang \orcidlink{0000-0002-6592-690X} \inst{1}
   \and
   Aditya M. Arabhavi \orcidlink{0000-0001-8407-4020} \inst{2}
   \and 
   Till Kaeufer \orcidlink{0000-0002-5430-1170} \inst{3}
   \and 
   Rens Waters \orcidlink{0000-0002-5462-9387} \inst{1,4}
   \and
   Inga Kamp \orcidlink{0000-0001-7455-5349} \inst{2}
   \and
   Thomas Henning \orcidlink{0000-0002-1493-300X} \inst{5}
   \and 
   Alessio Caratti o Garatti \orcidlink{0000-0001-8876-6614} \inst{6}
   \and 
   Ewine F. van Dishoeck \orcidlink{0000-0001-7591-1907} \inst{7,8}
   \and 
   Giulia Perotti \orcidlink{0000-0002-8545-6175} \inst{5,9} 
   \and 
   Jayatee Kanwar \orcidlink{0000-0003-0386-2178} \inst{10}
   \and
   Manuel G\"udel \orcidlink{0000-0001-9818-0588} \inst{11,12}
   \and
   Maria Morales-Calder\'on \orcidlink{0000-0001-9526-9499} \inst{13}
   \and
   Sierra L. Grant \orcidlink{0000-0002-4022-4899} \inst{14}   
   \and 
   Valentin Christiaens \orcidlink{0000-0002-0101-8814} \inst{15,16}   
   }
   
   \institute{Department of Astrophysics/IMAPP, Radboud University, PO Box 9010, 6500 GL Nijmegen, The Netherlands 
   \newline\email{hyerin.jang@astro.ru.nl}
   \and
   Kapteyn Astronomical Institute, Rijksuniversiteit Groningen, Postbus 800, 9700AV Groningen, The Netherlands 
   \and 
   Department of Physics and Astronomy, University of Exeter, Exeter EX4 4QL, UK 
   \and
   SRON Netherlands Institute for Space Research, Niels Bohrweg 4, NL-2333 CA Leiden, the Netherlands 
   \and  
   Max-Planck-Institut f\"{u}r Astronomie (MPIA), K\"{o}nigstuhl 17, 69117 Heidelberg, Germany 
   \and 
   INAF – Osservatorio Astronomico di Capodimonte, Salita Moiariello 16, 80131 Napoli, Italy 
   \and
   Leiden Observatory, Leiden University, PO Box 9513, 2300 RA Leiden, the Netherlands 
   \and
   Max-Planck Institut f\"{u}r Extraterrestrische Physik (MPE), Giessenbachstr. 1, 85748, Garching, Germany 
   \and
   Niels Bohr Institute, University of Copenhagen, NBB BA2, Jagtvej 155A, 2200 Copenhagen, Denmark 
   \and 
   Department of Astronomy, University of Michigan, 1085 South University Avenue, Ann Arbor, MI 48109, USA 
   \and
   Dept. of Astrophysics, University of Vienna, T\"urkenschanzstr. 17, A-1180 Vienna, Austria 
   \and
   ETH Z\"urich, Institute for Particle Physics and Astrophysics, Wolfgang-Pauli-Str. 27, 8093 Z\"urich, Switzerland 
    \and
   Centro de Astrobiolog\'ia (CAB), CSIC-INTA, ESAC Campus, Camino Bajo del Castillo s/n, 28692 Villanueva de la Ca\~nada, Madrid, Spain  
   \and
   Earth and Planets Laboratory, Carnegie Institution for Science, 5241 Broad Branch Road, NW, Washington, DC 20015, USA 
   \and 
   Institute of Astronomy, KU Leuven, Celestijnenlaan 200D, 3001 Leuven, Belgium 
   \and 
   STAR Institute, Universit\'e de Li\`ege, All\'ee du Six Ao\^ut 19c, 4000 Li\`ege, Belgium 
   }

   \date{\today}
    
\abstract
{
\textit{Context.}
Disks around very low-mass stars (VLMS) provide environments for the formation of Earth-like planets. Mid-infrared observations have revealed that these disks often exhibit weak silicate features and prominent hydrocarbon emissions.

\textit{Aims.}
This study aims to characterize the dust properties and geometrical structures of VLMS and brown dwarf (BD) disks, observed by the \textit{James Webb} Space Telescope (JWST)/Mid-Infrared Instrument (MIRI). We investigate how these properties relate to gas column density and potential evolutionary stages.

\textit{Methods.}
We analyze mid-infrared spectra of ten VLMS and BD disks from the JWST/MIRI observations as a part of the MIRI mid-Infrared Disk Survey (MINDS) program. Spectral slopes and silicate band strengths are measured and compared with hydrocarbon emission line ratios, which probe the gas column density. Moreover, the Dust Continuum Kit with Line emission from Gas (DuCKLinG) is used to quantify grain sizes, dust compositions, and crystallinity in the disk surface. 

\textit{Results.}
The disks are classified into less-, more-, and fully-settled geometries based on their mid-infrared spectral slopes and silicate band strengths. Less-settled disks show a relatively strong silicate band, high spectral slopes, and low crystallinity, and are dominated by 5~$\mu$m-sized grains. More-settled disks have weaker silicate band, low spectral slope, enhanced crystallinity, and higher mass fraction of smaller grains (<5 $\mu$m). Fully-settled disks exhibit little or no silicate emission and negative spectral slopes. An overall trend of increasing gas column density with decreasing spectral slope suggests that more molecular gas is exposed when the dust opacity decreases with increasing dust settling.

\textit{Conclusions.}
Our findings indicate that our sample shows dust processing signatures of grain growth and crystallization. These characteristics may reflect possible evolutionary pathways with disk turbulence, dust settling, and thermal processing or may alternatively point to inner-disk clearing or a collisional cascade. These results highlight the need for broader samples to understand the link between dust and gas appearance in regions where Earth-like planets form.  
}
\maketitle
\section{Introduction}
\label{sect:intro}
\paragraph{}
Disks around very low-mass stars (VLMSs), with stellar masses $\lesssim 0.3$ M$_{\odot}$ \citep{Liebert_Probst1987}, are compact with radii up to a few tens of au and have masses of the order of a few Jupiter masses \citep{Pascucci_etal2003,Klein_etal2003}. Observations show a high occurrence rate of $2.5\pm 0.2$ Earth to super-Earth planets on average around a VLMS in transit surveys \citep{Dressing_Charbonneau2015} and $1.32\pm 0.3$ planets in radial velocity studies \citep{Sabotta_etal2021}. These characteristics make VLMS disks ideal laboratories to investigate the formation of Earth-like planets within the inner disk. 

Dust is a crucial ingredient in the planet formation process. Among dust species, silicates are the most common in planet-forming disks as well as the interstellar medium \citep{Dorschner_etal1995, Colangeli_etal2003, Henning2010}. Micrometer-sized silicate dust absorbs stellar radiation, re-emits its energy, and features at mid-infrared wavelengths (mid-IR). In planet-forming disks, the mid-IR radiation originates from the disk surface above $\tau_{\rm dust} = 1$ at spatial distances that probe the terrestrial planet forming region. Thus, mid-IR silicate emission can probe the environments where Earth-like planets form. 

Silicate dust emits spectral features in the 10-70 $\mu$m wavelength range, and these features change with grain size, chemical composition, and lattice structure. Small grains ($0.1-1~\mu$m) produce strong and narrow features, while large grains ($\sim 5$ $\mu$m) have weaker and broader features \citep{Kessler_etal2006}. Thus, these features allow us to identify the grain sizes from spectral profiles. Moreover, different silicate species have distinct spectral shapes and peak positions, so an analysis of these features provides the compositional information of dust in the disks. The \textit{Spitzer} Space Telescope (\textit{Spitzer}) has previously observed these features in VLMS disks \citep{Apai_etal2005, Pascucci_etal2009}.

Studies have shown that dust in VLMS and brown dwarf (BD) disks is often processed into large and more crystalline grains, which suggests rapid grain growth. \cite{Apai_etal2005} investigated sub-stellar objects observed with \textit{Spitzer}. All those six BD disks have a broad 10 $\mu$m silicate band indicating large grains and 9-48 \% crystallinity in mass fraction. \cite{Pascucci_etal2009} confirmed that the 10 $\mu$m silicate band of disks around cool stars/BDs (M5–M9) is dominated by large and crystalline-rich dust grains compared to \textit{Spitzer} observations of Sun-like or Herbig Ae/Be stars. Molecular gas emissions are detected more frequently in flatter disks around Sun-like and cool stars, which is also in line with grain growth and dust settling. \cite{Kessler_etal2007} also showed larger grain sizes in BD disks compared to T Tauri and Herbig Ae/Be disks based on the strength and shape of the $10~\mu$m silicate band although the difference in crystallinity is marginal due to weaker luminosity of BDs. However, \textit{Spitzer} observations poorly constrain the molecular gas emissions around $6-8~\mu$m and $12-17~\mu$m regions. In particular, $6-8~\mu$m region exhibits a mixture of molecular gas features originating from the stellar photosphere and the disk, so it requires much higher resolution and sensitivity than \textit{Spitzer} to disentangle the sources of these features.

The \textit{James Webb} Space Telescope (JWST) has allowed more detailed investigations on VLMS and BD disks with its Mid-Infrared Instrument (MIRI). In this study, we refer these disks to VLMS disks, following \cite{Arabhavi_etal2025b} (hereafter referred to as Paper I). The M5.5 star Sz 28 is found to have a typical silicate grain size of 5 $\mu$m and crystallinities of the order of 20 \% \citep{Kanwar_etal2024} to 40 \% \citep{Kaeufer_etal2024}. On the other hand, J1605321-1933159 \citep{Tabone_etal2023} and ISO-Chal 147 \citep{Arabhavi_etal2024} do not show clear 10 and 20 $\mu$m silicate bands, possibly due to highly evolved dust populations dominated by grains larger than 5 $\mu$m. Such large dust grains lose their spectral features in the mid-IR. In addition, large dust grains settle toward the midplane, so VLMS disks are more settled compared to higher-mass stars \citep{Apai_etal2005}, which may affect the gas emission strength and thus detectability. 

In addition to dust, the spectral resolution and sensitivity of JWST/MIRI reveals rich molecular gas emission. Sz 28, J1605321-1933159, and ISO-Chal 147 exhibit strong hydrocarbon features, such as C$_2$H$_2$, C$_4$H$_2$, and C$_6$H$_6$, with weak or undetected H$_2$O and OH emissions \citep{Arabhavi_etal2025}. These results show that the three sources have a high C/O ratio in their inner disks \cite{Kanwar_etal2024}. On the other hand, Sz 114 \citep{Xie_etal2023} shows water-rich emission with C$_2$H$_2$ being the only hydrocarbon detected, and J04381486+2611399 \citep{Perotti_etal2025arXiv} also present only C$_2$H$_2$. These findings clearly demonstrate that VLMS disks exhibit a wide diversity in molecular gas compositions.

In Paper I, an overview is presented of the gas emissions detected from VLMS disks within the MIRI mid-Infrared Disk Survey \citep[MINDS; ][]{Henning_etal2024,Kamp_etal2023} JWST/MIRI Guaranteed Time Observations (GTO) program. They suggest a correlation between the dust and the gas in VLMS disks. Hydrocarbon-rich disks with strong C$_2$H$_2$, HCN, C$_4$H$_2$, and C$_6$H$_6$ have a weaker 10 $\mu$m silicate band and lower disk dust mass than oxygen-rich disks. 

In this paper, we perform a detailed analysis of the dust spectral features from the VLMS disks in the MINDS sample to identify dust properties, disk geometry, and their relation with gas column density. \Sect{MIRIspectra} introduces the sample and  describes their foreground extinction, silicate features, and overall spectral energy distributions (SEDs). We also analyze the relationship between spectral slopes and gas column densities measured in Paper I. In \sect{retrieval}, we use retrieval models to analyze silicate and hydrocarbon features. \Sect{discussion} discusses the modeling results, the link between spectral shapes and dust properties, and the presence of water absorption in stellar spectra. This paper is summarized in \sect{conclusion}.

\section{Mid-IR spectra of MINDS VLMS sample}
\label{sec:MIRIspectra}
\paragraph{}
In the MINDS program, ten disks around VLMSs were observed with JWST/MIRI Medium Resolution Spectroscopy (MRS) and reduced with the MINDS hybrid pipeline \citep{Christiaens_etal2024} , which leverages the standard JWST pipeline \citep[v1.14.1][]{Bushouse_etal2023}. Observations and the properties of the sample are described in Paper I, and we follow the source naming. 

The observed spectra of the ten VLMS disks are shown with gray lines in \fg{VLMSspectra}, and the spectra are rebinned to a lower spectral resolution of $R \sim 200$ (the black lines), which keeps the dust feature shapes but smears out the narrower gas emission lines. Among the ten sources, five sources (NC1, J1558, J0439, HKCha, and Sz28) exhibit clear 10 and 20 $\mu$m silicate emission bands. The remaining five other sources (NC9, IC147, TWA27, J1605, and J0438) show weak or no such features. 

Among the disks with the 10 $\mu$m silicate emission band, a range of crystalline silicate features is observed. J0439 shows the clearest crystalline bands over the MIRI spectrum. It has the distinct 10 $\mu$m enstatite feature and an 11 $\mu$m forsterite feature. At longer wavelengths, it has additional 16 and 20 $\mu$m forsterite features. In addition, NC9, HKCha, IC147, and J1605 show a weak bump at 21 $\mu$m, which we suspect to be related to the stoichiometry of SiO$_2$. 

However, not all disks have such well-defined crystalline signatures. Individual crystalline features are not visually distinguishable in NC1 and J1558, and many dust emission bands overlap each other \citep{Henning2010}. For instance, the 9 $\mu$m enstatite feature coincides with amorphous silica feature, and the 19 $\mu$m forsterite feature overlaps with the 19 $\mu$m enstatite feature. Thus, identifying dust species directly from the spectra is challenging. 

J0438 has a unique silicate feature around 10 $\mu$m. It has absorption at 9 $\mu$m and emission at 11 $\mu$m, which may be caused due to its high inclination $i =67$-$71^{\circ}$ \citep{Luhman2004,Scholz_etal2006}. Our study focuses on dust emission from less-inclined disks, which produce emissions over the disk. Thus, J0438 is excluded from this study, and a detailed study of its gas emission is discussed in \cite{Perotti_etal2025arXiv}. 

\begin{figure*}
    \centering
    \includegraphics[width=\linewidth]{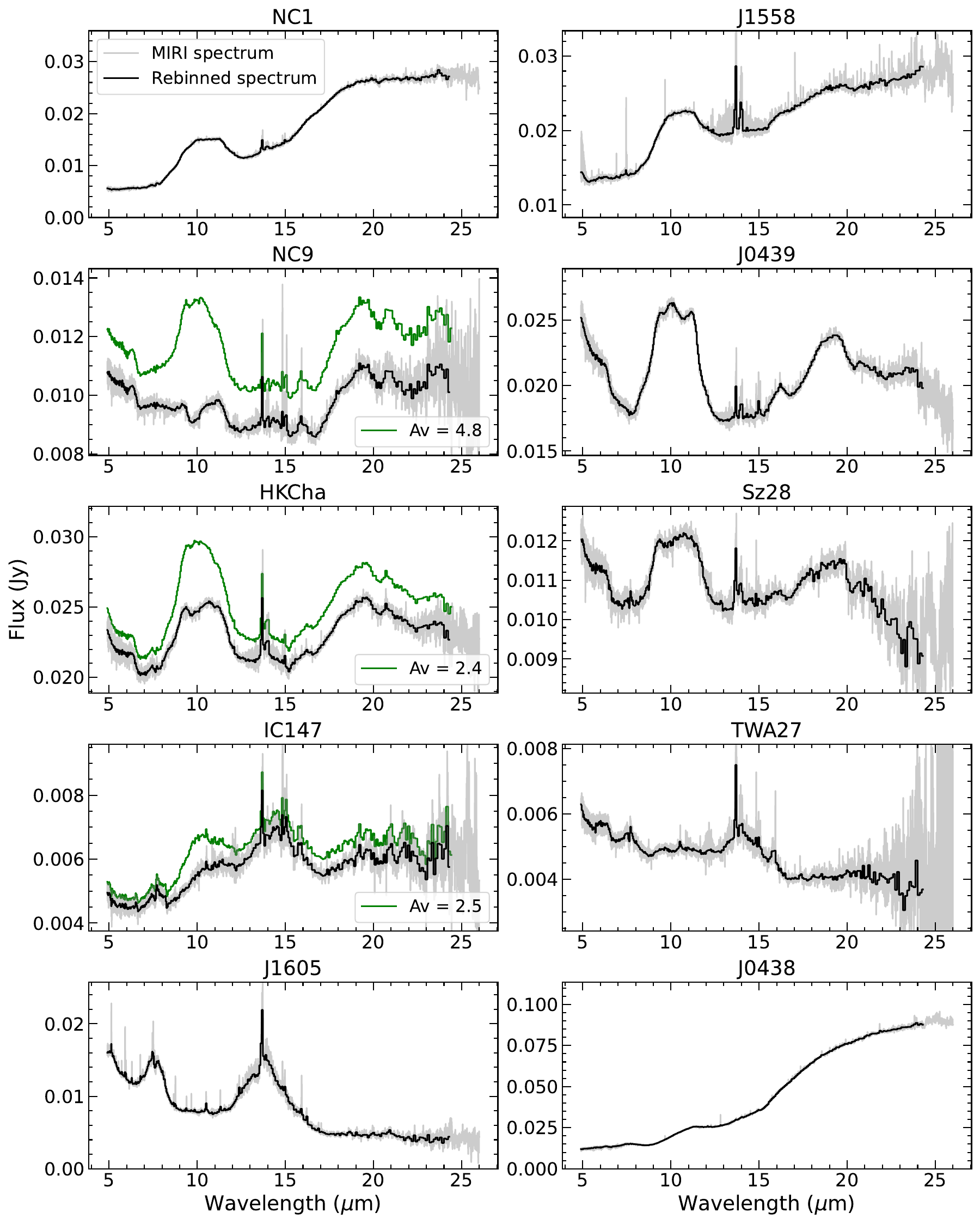}
    \caption{Spectra of VLMS disks in MINDS sample observed by JWST/MIRI MRS. The gray line is the MIRI spectrum, the black line is the rebinned spectrum to $R\sim200$. For NC9, HKCha, IC147, the green line is the foreground extinction corrected spectrum. The spectra are ordered based on their spectral slopes within the mid-IR range.}
    \label{fig:VLMSspectra}
\end{figure*}

\subsection{Foreground extinction}
\paragraph{}
The foreground extinction measurements are collected from \cite{Manara_etal2017,Herczeg_Hillenbrand2014,Herczeg_Hillenbrand2008} as described in Paper I. \cite{Manara_etal2017} model the observed spectra in optical wavelengths with a photospheric template and a slab model to reproduce the emission from accretion as in \cite{Herczeg_Hillenbrand2008}. They use a typical extinction law with $R_{\rm v} = 3.1$, following \cite{Cardelli_etal1989}. Most of our sample have low extinction ($A_{\rm v} < 2$) that has negligible impact on dust features in the mid-IR spectrum. For J1605 ($A_{\rm v} = 1.6$; the largest extinction below $A_{\rm v} = 2$), the extinction correction increases the flux around $10~\mu$m by only $7.6~\%$ while the flux of HKCha ($A_{\rm v}=2.4$; the smallest extinction above $A_{\rm v} = 2$ ) is increased by $12.7~\%$. In addition to the flux, the shape of the $10~\mu$m silicate band also changes as shown in \fg{VLMSspectra}. Moreover, typical uncertainties on extinction values for M-type stars studied in \cite{Manara_etal2017} are $\pm 0.5$ mag. Hence, the reliability of subtle variations in the mid-IR spectra remains challenging. Thus, we ignore the extinction correction for sources with $A_{\rm v} < 2$. 

NC9, HKCha, and IC147 have high foreground extinction ($A_{\rm v} > 2$), which suppresses the 10 $\mu$m silicate band. To recover the intrinsic silicate emission, we deredden the spectra with the extinction law prescribed in \cite{Gordon_etal2023,Gordon_etal2021,Gordon_etal2009,Fitzpatrick_etal2019,Decleir_etal2022} through the python package, \texttt{dust\_extinction} \citep{Gordon2024}. We use this extinction law because it covers the mid-IR range. The dereddened spectra are shown in green in \fg{VLMSspectra} and are used in the subsequent dust analysis. Before the extinction correction, NC9 already shows the 20 $\mu$m silicate band but not the 10 $\mu$m silicate band; the correction changes the shape of dust emission significantly. HKCha and IC147 show slightly stronger 10 $\mu$m silicate emission after the correction.

\subsection{Silicate dust emission}
\label{sec:10micronsilicate}
\paragraph{}
After correcting foreground extinction, we detect both the 10 $\mu$m and 20 $\mu$m silicate bands in six sources (NC1, J1558, NC9, J0439, HKCha, and Sz28). IC147 shows a weak 20 $\mu$m silicate band, but its 10 $\mu$m silicate band is not clearly identified due to the strong hydrocarbon emission. TWA27 and J1605 do not show silicate bands.

To quantify the 10 $\mu$m silicate band, we compared the band strength and shape in \fg{10micronBand} following \cite{vanBoekel_etal2003} and \cite{Olofsson_etal2009}. The band strength at 9.8 $\mu$m is $\text{F}9.8 = 1 + (f_{9.8\;\mu\rm m, cs} / <f_{\rm c}>)$, where $f_{9.8\;\mu\rm m,cs}$ is the linear continuum-subtracted flux at $9.8$ $\mu$m, and $<f_{\rm c}>$ is the average of the linear continuum from 7.7 $\mu$m to 12.7 $\mu$m. We follow the notations in \cite{Jang_etal2024a}. The shape of the 10 $\mu$m silicate band is characterized by the ratio F9.8/F11.3. F11.3 is the band strength at $11.3~\mu$m. This ratio becomes close to unity as the silicate band flattens and broadens due to grain growth and crystallization. Disks with large F9.8 tend to have smaller (micron or sub-micron) sized grains, while disks with small F9.8 have larger grains sizes of a few micron or more. In \fg{10micronBand}, our VLMS disks are located in the lower-left region, where the 10 $\mu$m silicate band is weak and broad compared to some other T Tauri disks and pre-transitional disks (gray dots) from the Combined Atlas of Sources
with \textit{Spitzer} IRS Spectra (CASSIS) database \citep{Lebouteiller_etal2011}. VLMS disks exhibit a narrower strength range that is approximately a factor of two lower than that of T Tauri disks. It indicates that the dust in the VLMS disks is more evolved and processed than that of the T Tauri disks. This trend has also been observed in VLMS disks using \textit{Spitzer} data \citep{Apai_etal2005, Pascucci_etal2009, Kessler_etal2007}, in comparison to T Tauri and Herbig disks. More recently, free-floating planetary-mass objects observed with NIRSpec and MIRI on JWST also exhibit weak and broad silicate emission bands, similar to those seen in VLMS disks and in contrast to the stronger features in T Tauri and Herbig disks \citep{Damian_etal2025}. A more detailed comparison between VLMS and T Tauri disks is provided in \cite{Grant_etal2025arXiv}.

\begin{figure}
    \centering
    \includegraphics[width=0.95\linewidth]{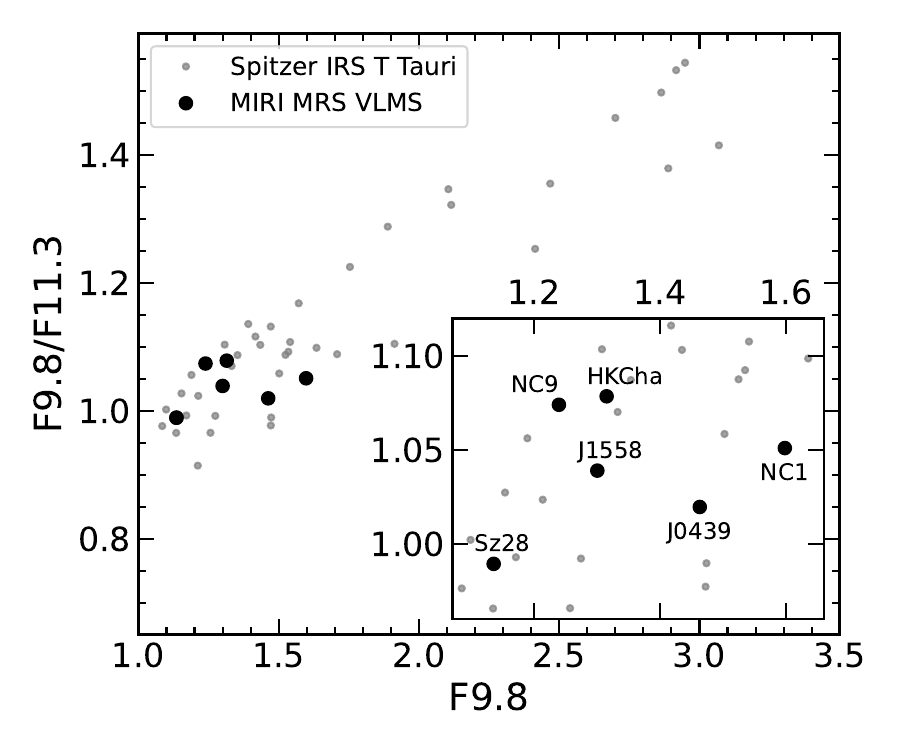}
    \caption{Strengths and shapes of the 10 $\mu$m silicate bands. The VLMS disks are shown in black dots, and gray dots are T Tauri disks observed with \textit{Spitzer} IRS extracted from the CASSIS database \citep{Lebouteiller_etal2011}. The small panel zooms into the region, where the VLMS disks are located.}
    \label{fig:10micronBand}
\end{figure}

\subsection{Mid-IR spectral slopes}
\label{}
\paragraph{}
The overall shapes of the VLMS mid-IR spectra differ across the sample. The flux levels of NC1 and J1558 increase with increasing wavelength, whereas NC9, J0439, HKCha, Sz28, and IC147 are relatively flat. TWA27 and J1605 show decreasing spectra. These slopes can be indicators of disk geometry \citep{Rilinger_Espaillat2021,Woitke_etal2016,DAlessio_etal2006}. \cite{Rilinger_Espaillat2021} model SEDs of BD disks with dust settling and show decreasing flux level from mid-IR to submillimeter wavelengths. \cite{Woitke_etal2016} present T Tauri disk models in which less-settled disks have increased flux at longer wavelength ($\lambda \gtrsim 20 ~\mu$m) compared to shorter wavelength ($\lambda \sim 10~\mu$m). In a less-settled disk, $\mu$m-sized dust grains efficiently emit in the disk surface above $\tau_{\rm mid-IR} = 1$ (vertically), including the colder outer disk (radially), emitting in the 20~$\mu$m region. On the other hand, a more-settled disk does not emit as much as the less-settled disk at longer wavelengths, so the flux is reduced. We note that the authors also mention the shape of the SED can be degenerate by multiple disk parameters, such as grain size distribution and disk structure. 

In Paper I, an anti-correlation between the $10~\mu$m silicate band strength and the hydrocarbon gas column density as measured from C$_2$H$_2$ and $^{13}$CCH$_2$ are presented. It suggests that the $\tau_{\rm mid-IR} = 1$ surface moves closer to the midplane with increasing dust settling and exposes a larger gas column. In this study, we further investigate this scenario, using the spectral slopes as a probe of optical depth and gas column density. 

We measure the spectral slope between $\lambda_{1}=12.6 \;\mu\rm m$ and $\lambda_{2} = 22.5 \;\mu\rm m$, where silicate dust features are weak,
\begin{equation}
    \alpha = \dfrac{\log_{10}F_{\nu}(\lambda_{1})- \log_{10}F_{\nu}(\lambda_{2})}{\log_{10}\lambda_{1} - \log_{10}\lambda_{2}}.
\end{equation}
For J1605, we set $\lambda_{1}=11.5~\mu\rm m$ to avoid the strong hydrocarbon emission. 

In our sample, the spectral slopes vary from -1 for J1605 to 1.5 for NC1. We compare this spectral slope to the 10 $\mu$m silicate band strength (F9.8) in \fg{F98slope}. Since J1605, TWA27, and IC147 do not have a clear 10 $\mu$m silicate band, we use $\rm F9.8 = 1$ for these disks as the lower limit. The silicate band in IC 147, identified in \sect{duckling_results}, is measured to have F9.8 = 1.23, as indicated by the gray dotted arrow in \fg{F98slope}. In \fg{F98slope}, the silicate band strength gets stronger with increasing spectral slope. The high spectral slope can be interpreted as more flux from the outer disk, which emits at longer wavelengths. The strong silicate band strength can be due to more smaller grains ($0.1-1~\mu$m) than larger grains ($4-5~\mu$m) or large abundance of $\mu$m-sized grains ($0.1-5~\mu$m) in the disk surface. Whether the disk surface is rich in small grains ($\lesssim 1~\mu$m) or in $\mu$m-sized grains (sub to a few $\mu$m), both represent the disk is less-settled because these grains are efficient to vertically stir up. Thus, in this paper, we refer to a disk with high spectral slope as a less-settled disk and low spectral slope as a more-settled disk.

\begin{figure}
    \centering
    \includegraphics[width=0.95\linewidth]{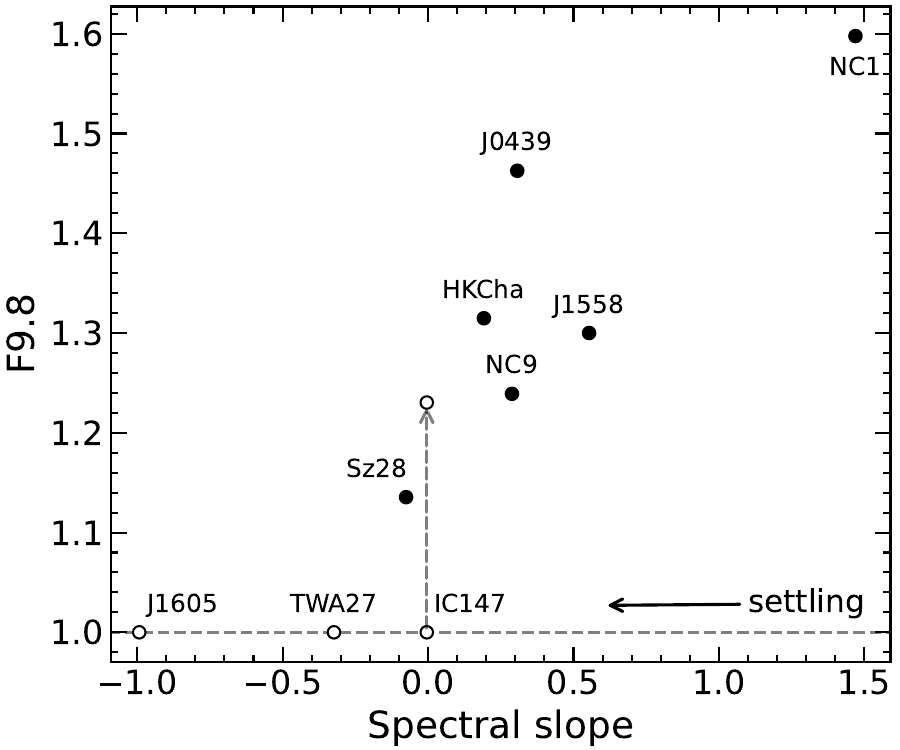}
    \caption{Strength of the 10 $\mu$m silicate band and $12.6-22.5~\mu$m spectral slope of the VLMS disks. The silicate bands for J1605, TWA27, IC147 are set to be unity due to the absence of a clear silicate band. The F9.8 value of IC147 from its decomposed disk components in the DuCKLinG model is F9.8=1.23. }
    \label{fig:F98slope}
\end{figure}

We further compare the spectral slope with the integrated flux ratio of C$_2$H$_2$ to $^{13}$CCH$_2$, measured in Paper I, in \fg{13CCH2C2H2slope}. The flux ratio ($F_{^{13}\rm CCH_2}/F_{\rm C_{2}H_{2}}$) serves as a proxy for gas column density because C$_2$H$_2$ becomes optically thick and saturates at high column densities, while less abundant $^{13}$CCH$_2$ stays optically thin up to higher column densities and continues to increase in flux. This different saturation allows the ratio to reflect variations in column density \citep{Arabhavi_etal2025b}. In \fg{13CCH2C2H2slope}, the lower-left region stays empty, which indicates more-settled disks do not show low gas column density. Instead, more-settled disks tend to have higher gas column density. Because a more-settled disk has fewer dust grains providing opacity in the disk surface than a less-settled disk, the $\tau_{\rm mid-IR}=1$ surface gets deeper into the disk. Thus, it is possible to observe gas molecules deeper towards midplane, which should result in high gas column density with large $F_{^{13}\rm CCH_2}/F_{\rm C_{2}H_{2}}$ value. This is consistent with the anti-correlation between the gas column density and dust strength reported in Paper I. As the integrated flux of the $10~\mu$m silicate band becomes larger, $F_{^{13}\rm CCH_2}/F_{\rm C_{2}H_{2}}$ gets smaller. Thus, Paper I also suggests that disks with a higher dust opacity show hydrocarbon-poor spectra.

\begin{figure}
    \centering
    \includegraphics[width=0.95\linewidth]{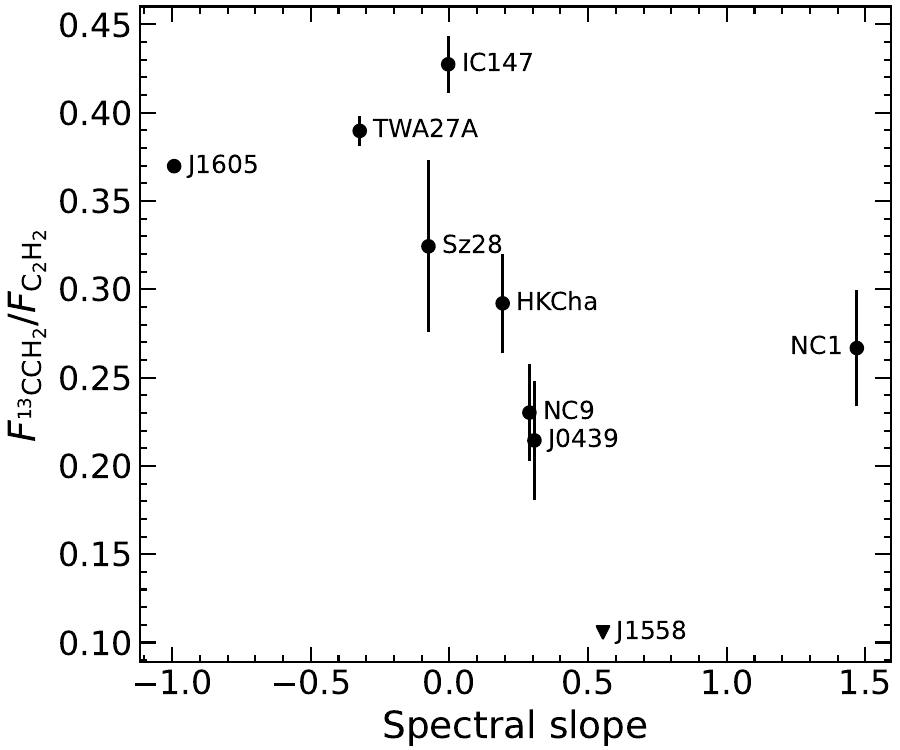}
    \caption{Gas column density and spectral slope. $F_{^{13}\rm CCH_{2}}/F_{\rm C_{2}H_{2}}$ represents gas column density, measured in Paper I. J1558 is the upper limit value for $F_{^{13}\rm CCH_{2}}/F_{\rm C_{2}H_{2}}$.}
    \label{fig:13CCH2C2H2slope}
\end{figure}

\subsection{Stellar spectra}
\label{sec:stellarspectra}
\paragraph{}
While the mid-IR spectrum at longer wavelengths traces emission from the cooler dust in the outer disk, shorter wavelengths below 8 $\mu$m capture the inner disk contribution as well as the stellar photosphere. In our sample, five disks (NC9, J0439, HKCha, Sz28, and TWA27) show a step-like feature at 6.5 $\mu$m. Specifically, the flux decreases sharply from 6.35 $\mu$m to 6.65 $\mu$m. This step-like feature does not correspond to known dust or gas emission bands from the disk but matches with a water absorption feature seen in stellar photospheres (\fg{J0439water}). In contrast, four other disks (NC1, J1558, IC147, J1605) do not have this feature in the MIRI spectra. This suggests that their inner disk emission is dominant enough to minimize the visible signature of the underlying stellar water absorption at 6.5 $\mu$m, or the central star already has very weak water absorption at 6.5~$\mu$m. The presence or absence of this absorption feature in the MIRI spectra provides a useful diagnostic to measure the inner disk contribution on top of the stellar photosphere. Thus, the contribution of the stellar photosphere in mid-IR should also be studied to analyze MIRI spectra of the VLMS disks.

We model stellar photospheric spectra for each source using the VO SED Analyzer \citep[VOSA; ][]{Bayo_etal2008}. VOSA collects photometric data from multiple catalogs and provides a best-fit model to the observations using grids of theoretical models. For each source, we compiled photometry from the \textit{Gaia} Data Release 3 (DR3), the Two Micron All Sky Survey (2MASS), and the Wide-field Infrared Survey Explorer (WISE) to fit BT-Settl models \citep{Allard_etal2003, Allard_etal2007,Allard_etal2011,Allard_etal2012,Allard_etal2013,Barber_etal2006, Caffau_etal2011} using chi-square fits. The fitted stellar photospheric spectra and their photometric data are shown in \fg{SED}, and the resulting effective temperature, gravity, bolometric luminosity are listed in Table \ref{tab:VOSA}. For NC9, the \textit{Gaia} DR3 Gbp and G optical photometric data were excluded from the fit; the high extinction ($A_V$) affects the optical data, which led to poorer fits in the IR. All photometric data were dereddened based on the collected extinction values in Paper I and shown in \fg{SED}. 

Overall, the stellar and MIRI spectra align well with photometric data. An exception is J1605, which has strong and variable IR excess around 5 $\mu$m \citep{Tabone_etal2023}; this causes a mismatch between WISE and MIRI observations around 5 $\mu$m. Nevertheless, the rest of its photometry is well-fitted, so we use this stellar model for the measurements of the water absorption feature in \sect{W_model}. The modeled photometric spectra are also provided to mid-IR retrieval models for further analysis in the following section.

\section{Mid-IR retrieval model}
\label{sec:retrieval}
\paragraph{}
Due to the ambiguity and blending of dust features with molecular gas emission, we perform a consistent retrieval analysis to constrain the mineralogy of dust in the VLMS disks. In the analysis, we decompose the MIRI spectra into multiple disk components, including inner rim, midplane, disk surface, and gas emission by using the Dust Continuum Kit with Line emission from Gas (DuCKLinG; \cite{Kaeufer_etal2024a}).

We used DuCKLinG to identify and characterize the dust components in the MIRI spectra. DuCKLinG employs the two-layer disk model introduced by \citet{Juhasz_etal2009, Juhasz_etal2010}, which is not suitable for highly inclined disks such as J0438. All other targets in our sample have inclinations less than $45\degree$. We compute the posterior distribution using Bayesian nested sampling via MultiNest \citep{Feroz_Hobson2008,Feroz_etal2009,Feroz_etal2019}, accessed through the Python interface PyMultiNest \citep{Buchner_etal2014}. A standard Gaussian likelihood function is adopted with uniform priors for all parameters \citep[following][]{Kaeufer_etal2024a}. We let the retrievals run with an evidence tolerance of 5 and extract one representative model (the median probability model) for visual comparisons to the observed data. DuCKLinG simultaneously models the gas and dust contributions in the mid-IR. This joint approach is necessary because our VLMS disks have significant pseudo-continuum emission from hydrocarbons, which makes it difficult to distinguish dust features from the gas emission. Since the focus of this study is on dust mineralogy, we include gas emission modeling at a level to distinguish between the gas and dust features. A more detailed analysis of the gas emission features is in Paper I. Therefore, we abstain from conclusive interpretations of the gas components that we retrieve.

\subsection{DuCKLinG model setup}
To focus on the dust features, we degrade the resolution of the MIRI spectra from $R \sim 1500-3000$ to a constant spectral resolving power of $R \sim 200$, comparable to the \textit{Spitzer} IRS resolution. This rebinning is performed with the python package \texttt{coronagraph} \cite{Robinson_etal2016,Lustig_etal2019}. Each data point is weighted by its expected signal-to-noise ratio, calculated from the JWST Exposure Time Calculator, following the methodology of \cite{Temmink_etal2024}. The Long sub-band of Channel 4 ($24.40-27.90~\mu$m) is neglected in this study due to its significantly low signal-to-noise ratio. 

In the wavelength range from 11.7 $\mu$m to 16.8 $\mu$m, we keep the full spectral resolution of the MIRI spectra to allow DuCKLinG to more accurately fit the emissions of hydrocarbons. If low spectral resolution is used in this region, the retrieval results may misinterpret molecular gas emissions (as discussed in \sect{R200}). Therefore, it is important to maintain high resolution in regions rich in gas emission features. The data points in this high-resolution region are assigned lower weights to prevent overfitting of gas features with poor fitting for the dust features. This approach balances the influence of gas and dust in the retrieval model and ensures that the dust analysis remains the primary focus. \Sect{R200} compares the retrieved results of fully $R \sim 200$ spectrum of Sz28 and detailed gas study of Sz28 in \cite{Kaeufer_etal2024}.

We adopt dust opacities generated from the Gaussian random field model \citep[GRF; ][]{Min_etal2007}, as these provide the best fit to JWST/MIRI MRS observation for the PDS 70 disk in \cite{Jang_etal2024b}. Our analysis focuses on five common dust species found in the interstellar medium and protoplanetary disks. Two are crystalline silicates: forsterite \citep{Servoin_Piriou1973} and enstatite \citep{Jaeger_etal1998}. The remaining three are amorphous: Mg$_2$SiO$_4$ \citep{Henning_Stognienko1996}, MgSiO$_3$ \citep{Dorschner_etal1995}, and SiO$_2$ \citep[silica; ][]{Henning_Mutschke1997}. Each species has six grain sizes of 0.1, 1, 2, 3, 4, 5 $\mu$m. Grains larger than 5 $\mu$m do not show clear spectral features in the mid-IR and are excluded. In addition, featureless dust species in the mid-IR, such as metallic iron and amorphous carbon, cannot be investigated with this approach.

We include common gas-phase molecules with strong mid-IR emission around 14 $\mu$m, such as C$_2$H$_2$, C$_6$H$_6$, HCN, CO$_2$, and C$_4$H$_2$, based on their detection reported in Paper I. These molecular features are modeled with a single emitting temperature per species, which simplifies the fitting process and minimizes computational cost while maintaining sufficient accuracy for our dust-focused study.

\subsection{DuCKLinG results}
\label{sec:duckling_results}
\paragraph{}
\Fgs{fit_13}{fit_17} present the final retrieval models with residuals. \Fg{mass_13} shows the mass fractions of dust species in the disk surface. Values of mass fractions with error bars are summarized in Table \ref{tab:mass_frac}.

The final models (black line) reproduce the observed spectra (orange line) with residuals below 10 \% for dust features. However, optically thin gas emissions, which appear as sharp peaks, show residuals exceeding 10 \% due to the simplified gas fitting. Given our focus on the overall dust features in the VLMS disks, detailed gas features are not analyzed further. In this section, we describe the retrieval results for each VLMS disk.

\begin{figure*}
    \centering
    \includegraphics[width=\textwidth]{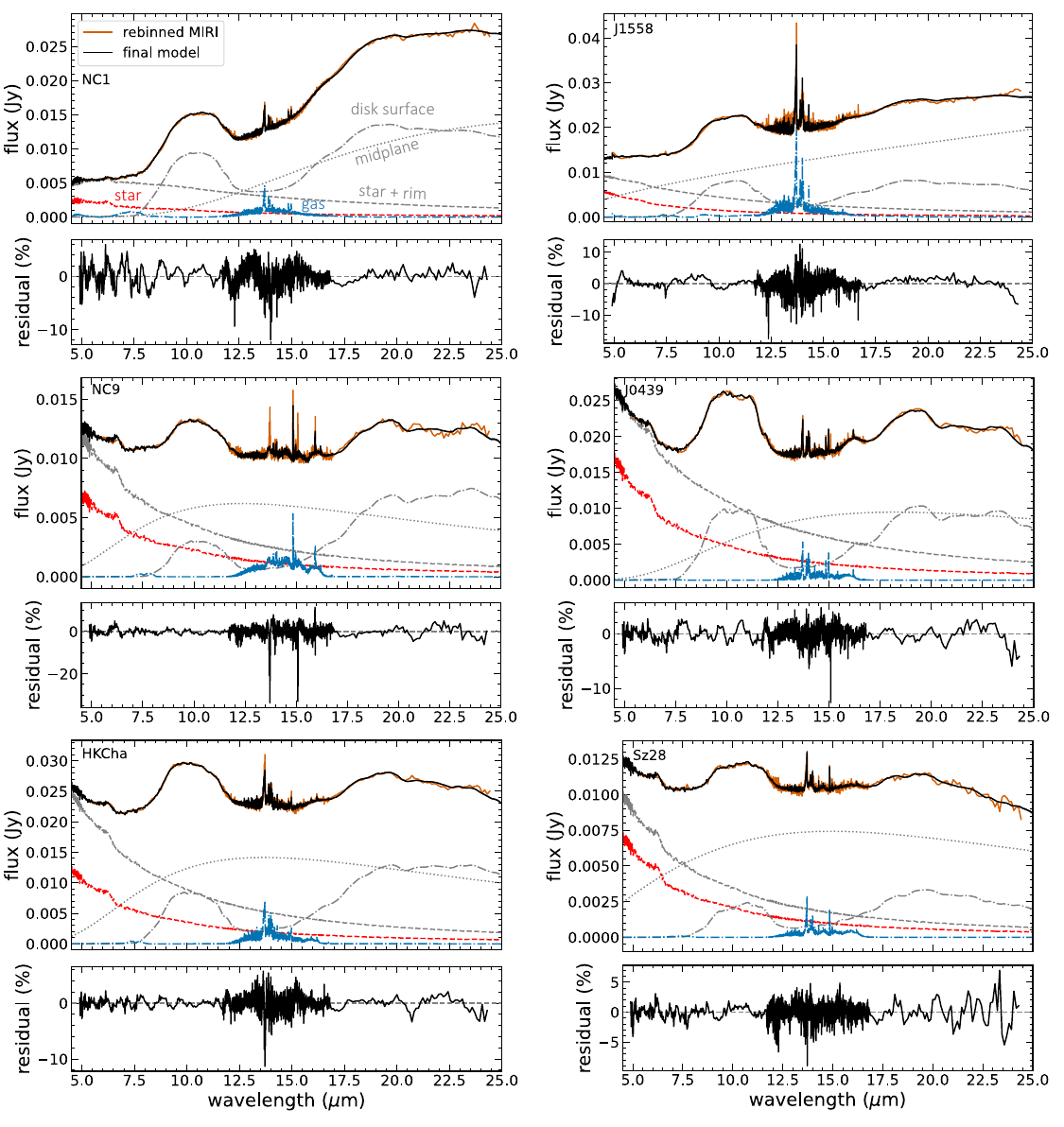}
    \caption{DuCKLinG fitting results for NC1, J1558, NC9, J0439, HKCha, and Sz28. The final model is the black solid line, and the rebinned MIRI spectrum is the orange line. The gray dashed, dotted, and dash-dotted lines are inner rim and stellar (red line) combined component, the midplane component, and the disk surface component, respectively. The blue line is the gas emission component.}
    \label{fig:fit_13}
\end{figure*}

\begin{figure}
    \centering
    \includegraphics[width=\linewidth]{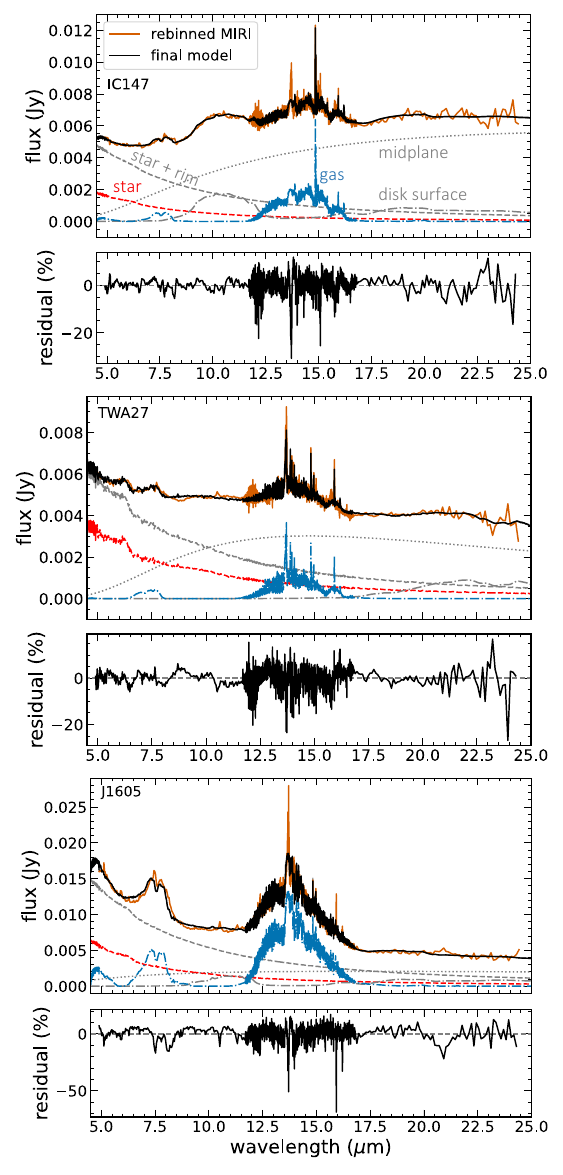}
    \caption{DuCKLinG fitting results for IC147, TWA27, and J1605. Each line represents as same as in \fg{fit_13}. }
    \label{fig:fit_17}
\end{figure}

\begin{figure*}
    \centering
    \includegraphics[width=0.95\textwidth]{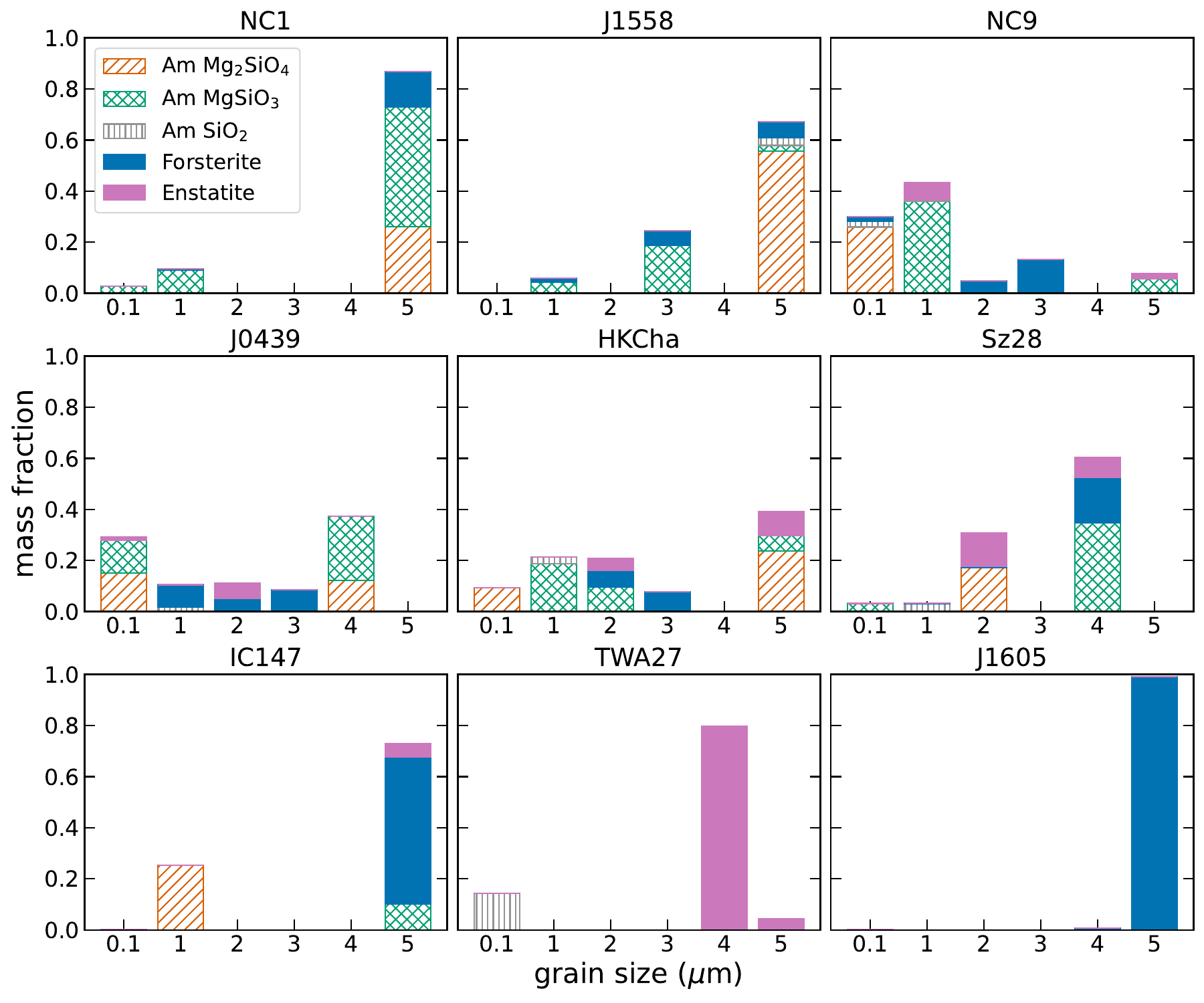}
    \caption{Mass fractions of dust species and grain sizes resulting from the DuCKLinG models.  }
    \label{fig:mass_13}
\end{figure*}

NC1: The final model reproduces the observed spectrum with overall median residuals of 1.1 \%. The overall residual represents the median of the absolute residuals. $5~\mu$m-sized forsterite is the only detected crystalline silicates, while 5 $\mu$m-sized amorphous Mg$_2$SiO$_4$ and MgSiO$_3$ dominate the mass fraction.

J1558: The model agrees with the observation within overall median 1.3 \% residuals. Detected dust species are mainly 5 $\mu$m-sized amorphous Mg$_2$SiO$_4$ and 3 $\mu$m-sized amorphous MgSiO$_3$. Other dust species contribute less than 10 \% in mass, and enstatite is not detected. 

NC9: The model fits the observed spectrum well with overall median residuals 1.3 \%, but the 21 $\mu$m feature is not well fitted by our model. The spectrum is dominated by small amorphous silicates (0.1 and 1 $\mu$m), while forsterite grains are relatively large (3 $\mu$m).

J0439: The model provides a good fit, with overall median residuals 0.8 \%. Both small (0.1\,$\mu$m) and large (4\,$\mu$m) amorphous silicates are detected. Forsterite and enstatite are found in intermediate grain sizes (1--3 $\mu$m). Forsterite is more abundant in smaller sizes compared to enstatite, which is expected from the strong forsterite features in the 16--23 $\mu$m region. 

HKCha: The model reproduces the spectrum with overall median  residuals 0.8 \%. Dominant components are 0.1 and 5\,$\mu$m amorphous Mg$_2$SiO$_4$ and 1 and 2\,$\mu$m MgSiO$_3$. Forsterite grains are 2 and 3\,$\mu$m in size, while enstatite grains are 2 and 5\,$\mu$m. It also has the the unfitted 21 $\mu$m feature. 

Sz28: Dust features are well fitted with overall median  residuals 0.8 \%. Dust components include 2\,$\mu$m-sized amorphous Mg$_2$SiO$_4$, 4\,$\mu$m-sized amorphous MgSiO$_3$, 4\,$\mu$m-sized forsterite, and 2\,$\mu$m-sized enstatite. 

IC147: The 10\,$\mu$m region is well reproduced with 5\,$\mu$m-sized forsterite, 5\,$\mu$m-sized amorphous MgSiO$_3$, and 1\,$\mu$m-sized amorphous Mg$_2$SiO$_4$. We note that this result differs from that in \cite{Arabhavi_etal2024} because in that paper no correction for foreground extinction was taken into account. The overall median residual is 2.2 \%.

TWA27: Despite its flat spectrum, the model indicates contributions from large grains such as $4~\mu$m-sized enstatite. The $10~\mu$m region is dominated by cold midplane emission with no clear dust features. However, emission at longer wavelengths still requires the silicate dust to explain the observed flux. The overall median residuals are 2.7 \%.

J1605: The spectrum is flat with dominant hydrocarbon emissions and innermost disk contribution. However, 5\,$\mu$m-sized forsterite is still required to reproduce the slight excess near the 10 and 20\,$\mu$m silicate bands. Residuals are 4 \%, and the 21 $\mu$m feature is not reproduced by our model.

Although IC147, TWA27, and J1605 do not have clear 10 $\mu$m and 20 $\mu$m silicate bands, the DuCKLinG models suggest the presence of silicate dust in all three sources. To validate the silicate detections in these spectra, we include both flat opacity and C$_2$H$_4$ gas into the DuCKLinG model. C$_2$H$_4$ produces a broad pseudo-continuum around 10.5~$\mu$m, so we use the full spectral resolution of the MIRI data from 8.26 $\mu$m to 16.8 $\mu$m to fit C$_2$H$_4$  emission. A flat opacity is an artificial opacity with unity absorption efficiency across the mid-IR. In DuCKLinG, this opacity shows up as a blackbody spectrum with the temperature of the disk surface and represents any featureless dust species that may contribute to the continuum in mid-IR. For quantitative calculation, we set the grain size of this opacity to 5 $\mu$m, and their retrieved mass fractions are summarized in Table \ref{tab:mass_frac2}. 

In IC147, C$_2$H$_4$ emission only contributes $2.5~\%$ around $10~\mu$m although it provides a significant contribution in \cite{Arabhavi_etal2024}. This is because the dust continuum is fitted here to the extinction corrected IC147 spectrum. Mass fractions of 5~$\mu$m-sized silicates are significantly replaced by flat opacity, but the 0.1 $\mu$m-sized amorphous Mg$_2$SiO$_4$ still produces a distinct 10 $\mu$m silicate band and a weak 20 $\mu$m bump. In TWA27, the final model still produces the cold silicate dust emission at longer wavelengths as seen in \fg{fit_13}. In J1605, the disk surface contributes to the final model only $\sim 0.05$ \%. This suggests that the J1605 spectrum can be modeled without the disk surface component. To conclude, silicate dust grains are present in the disk surfaces of IC147 and TWA27, but their detailed composition is uncertain due to the weak spectral features. Notably, the presence of silicate dust in J1605 is unreliable. It remains the possibility that the strong hydrocarbon emissions completely overwhelm the silicate emission, but this cannot be assessed within this modeling framework. 

NC9, HKCha, and J1605 show an unidentified feature near 21 $\mu$m that is not reproduced by the available dust opacities in DuCKLinG. Although this feature resembles silica emission at 21 $\mu$m, the absence of the accompanying prominent 9 $\mu$m silica feature makes the detection of silica uncertain. The residual at 21 $\mu$m is not strong enough to affect the overall fit, but it points to the need for exploring additional dust species or alternative optical constants to investigate this undefined feature. The quality of data now warrants to revisit how to measure and model the dust opacities.

\section{Discussion}
\label{sec:discussion}
\paragraph{}
In this section, we compare our results to previous studies of individual sources. We also discuss the stellar water absorption feature with the DuCKLinG results, classify the VLMS disks into three categories, and examine disk crystallinity.

\subsection{Individual studies}
\paragraph{}
Several individual sources (J1605,  IC147, Sz28, and NC1) have been previously studied in detail with a focus on gas emission features. \cite{Tabone_etal2023} analyze the two strong hydrocarbon emissions at 7.7 $\mu$m and 13.7 $\mu$m for J1605 and suggest two-component model of C$_2$H$_2$ with optically thin and thick emissions. The optically thin component reproduces the sharp C$_2$H$_2$ peak, while the optically thick component contributes to a broad pseudo-continuum. Our model does not adopt this two-component C$_2$H$_2$ framework, so the sharp peak is poorly fitted. This could be a reason that multiple peaks of gas emission appear in the residuals of our DuCKLinG model. 

\cite{Arabhavi_etal2024} identify 13 hydrocarbons emitting in IC147. These hydrocarbon emissions are significant enough to analyze the spectrum without silicate dust emission. In our analysis, we consider the foreground extinction, which affects the 10 $\mu$m silicate dust emission. After correcting the extinction, silicate dust emission from forsterite and amorphous Mg$_2$SiO$_4$ are detected. Before the retrieval modeling, the silicate feature in IC147 was not clearly visible, and its F9.8 value was set to 1 due to the lack of measurable emission in the MIRI spectrum. From the retrieved model, we can isolate the silicate feature and measure the F9.8 value. From the modeled spectrum, we subtracted the gas emission component and normalized the 10 $\mu$m silicate band as described in \sect{10micronsilicate}. The new F9.8 value increases to 1.23 and follows the trend of the F9.8-slope correlation shown in \fg{F98slope}.

For Sz28, \cite{Kanwar_etal2024} investigate gas-phase chemistry with C/O > 1 using thermo-chemical models, and it was found to be consistent with the detection of hydrocarbons and the non-detection of oxygen-bearing molecules. They used DuCK, a dust-only version of DuCKLinG, to estimate the dust continuum and report typical grain sizes of 2-5 $\mu$m with a crystallinity around 20 \%. \cite{Kaeufer_etal2024} extended the analysis with DuCKLinG and found dominant 5 $\mu$m-sized grains with 44 \% crystallinity. We also find in our analysis a crystallinity of $\sim40~\%$, consistent with \cite{Kaeufer_etal2024}. Half the crystallinity of \cite{Kanwar_etal2024} can be due to the fact that they divided the spectrum into short ($\lambda < 15~\mu$m) and long ($\lambda > 15~\mu$m) wavelength regions. We also note that \cite{Kaeufer_etal2024} use a more limited set of grain sizes ($0.1~\mu\rm m,\; 2~\mu\rm m$, and $5~\mu$m), whereas our model has a finer grain size distribution.

\cite{Morales_Calder_etal2025arXiv} characterize the gas and dust composition of the NC1 disk and find various hydrocarbons, which indicate C/O > 1, and dominant $4~\mu$m-sized amorphous MgSiO$_3$ with 13 \% of $5~\mu$m-sized forsterite in mass fraction. In this study, we find dominant $5~\mu$m-sized amorphous MgSiO$_3$. It can be due to the subtle difference in opacity between $4~\mu$m and $5~\mu$m grain sizes. We also detect 14 \% of $5~\mu$m-sized forsterite, and enstatite is not detected in either studies.

\subsection{Stellar water absorption feature}
\label{sec:W_model}
\paragraph{}
Some of MIRI spectra of the VLMS disks show a step-like feature at $6.5~\mu$m. It naturally leads us to wonder about the nature of the feature, which could easily be mistaken for dust features. At 6.5 $\mu$m, all stellar models exhibit the water absorption feature, but only five VLMS disks show the step-like feature in the MIRI spectra. We postulate that the absorption feature is diluted by the IR excess from the disk. We note that the photospheric water absorption feature may still be detectable even with the substantial disk contribution if the signal-to-noise of the data is high enough.

We quantify the water absorption feature by introducing its depth in both the modeled stellar ($W_{\rm star}$) and MIRI ($W_{\rm obs}$) spectra. We define $W$ as the difference in flux measured at 6.35~$\mu$m ($F_{6.35 \mu m}$) and 6.65~$\mu$m ($F_{6.65 \mu m}$): $W = F_{6.35 \mu m}-F_{6.65 \mu m}.$ \Fg{waterdeep} shows that the depth of absorption tends to be deeper in the MIRI spectra if it is deeper in the stellar spectra, which supports that the step-like features in MIRI spectra originate from the stellar photosphere water absorption feature.

If a disk contributes no emission in this region, the depth measured from MIRI spectra should be the same as in the stellar spectra ($W_{\rm star} = W_{\rm obs}$). If the disk emission adds equal amounts of flux at 6.35 and 6.65 $\mu$m, we also expect $W_{\rm star} = W_{\rm obs}$ even in the case of a substantial disk contribution. Inspection of \fg{waterdeep} illustrates that the MIRI spectra consistently show shallower absorption. Especially, the NC1 spectrum does not show the absorption feature ($W_{\rm obs} =0$), and the J1605 spectrum even has a higher flux at 6.65 $\mu$m ($W_{\rm obs} < 0$). It requires strong disk contribution around 6.5 $\mu$m to completely obscure the absorption for NC1 and J1605. In J1605, it is clear from the spectrum that the hydrocarbon emissions overwhelm the absorption feature. The absorption feature in NC1 may be also obscured by gas emission. We note that this analysis relies on the accuracy of the stellar models used, and any inaccuracies in the models can lead to deviations.

\begin{figure}
    \centering
    \includegraphics[width=0.95\linewidth]{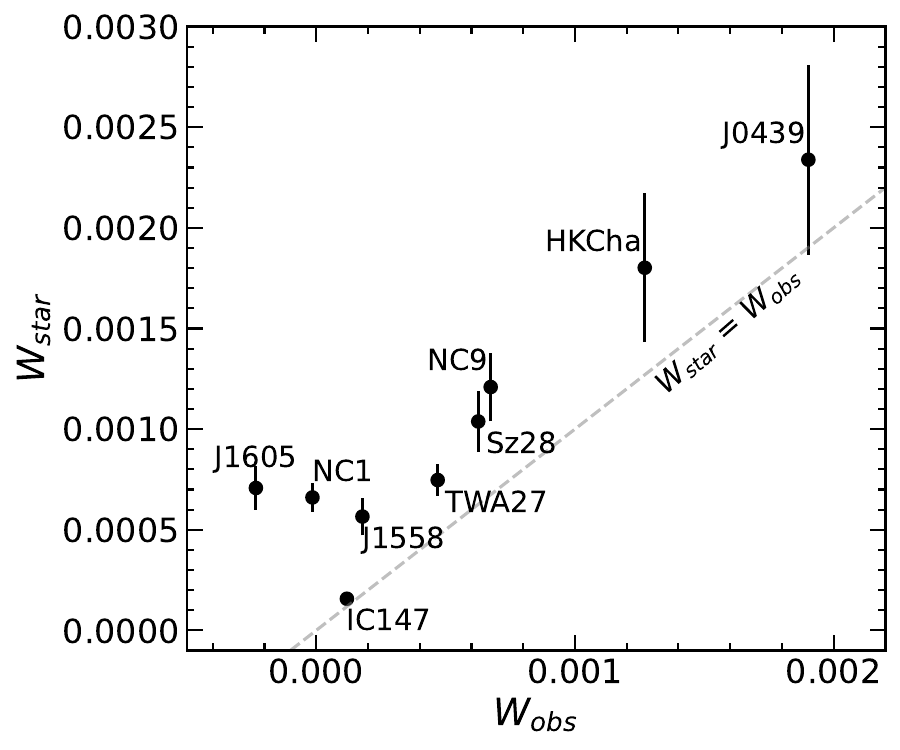}
    \caption{Depths of water absorption feature in stellar and MIRI spectra. The dashed line represents the depth of water absorption in the MIRI spectrum being the same as that in the stellar spectra. The error bars (black vertical lines) are estimated based on the uncertainties in the total fluxes of the stellar models. }
    \label{fig:waterdeep}
\end{figure}

We decompose the MIRI spectra with DuCKLinG to examine which disk component contributes to the suppression of this absorption feature. \Fg{water_diskcomp} presents the contribution of each spectral component to the observed depth of the water absorption feature. The steep blackbody slope of the inner rim generally deepens the depth of the stellar water absorption feature except in the case of NC1, where the inner rim instead acts as a positive component and reduces the depth. For all sources, midplane, gas emission, and disk surface components diminish the water absorption feature. We note that the sum of contributions on each source does not exactly reproduce the observed depth because these are the retrieval-model-based measurements. The retrieval models have lower resolution than the observation and are not perfectly reproducing the observation. 

Gas emission significantly reduces the depth of the water absorption feature in J1605, followed by NC1. The midplane component generally weakens the absorption as well except in IC147 and NC1, and only marginally in J1605. This indicates that gas emission is mainly responsible for the blending of stellar water absorption in NC1 and J1605. Additional emission from warm water emission around 6.8 $\mu$m may also contribute to the weakening of the absorption feature. Warm water emission from these disks are reported in \cite{Arabhavi_etal2025} and \cite{Morales_Calder_etal2025arXiv}. 

In IC147, the disk surface component significantly contributes to the weakening of the absorption feature through the flat opacity. In HKCha, the midplane component substantially reduces the depth of the absorption feature. However, the feature is strong in the stellar spectrum that it remains visible in the MIRI spectrum. 

The absent or very weak water absorption feature in both NC1 and J1558 arises from different origins. Gas emission and midplane components are responsible in NC1 and J1558, respectively. We note that DuCKLinG does not account for all possible gas emission in the mid-IR range. Therefore, unidentified gas emission may further contribute to the observed suppression of stellar water absorption features, in addition to the modeled midplane component.

In conclusion, our decomposition of the MIRI spectra demonstrates that the observed suppression of the stellar water absorption feature can be primarily attributed to excess emission from the disk, either from the optically thick midplane or from hot gas in the inner disk, depending on the source. These results highlight the significant role of disk emission in shaping mid-IR spectral features in observation.

\begin{figure}
    \centering
    \includegraphics[width=\linewidth]{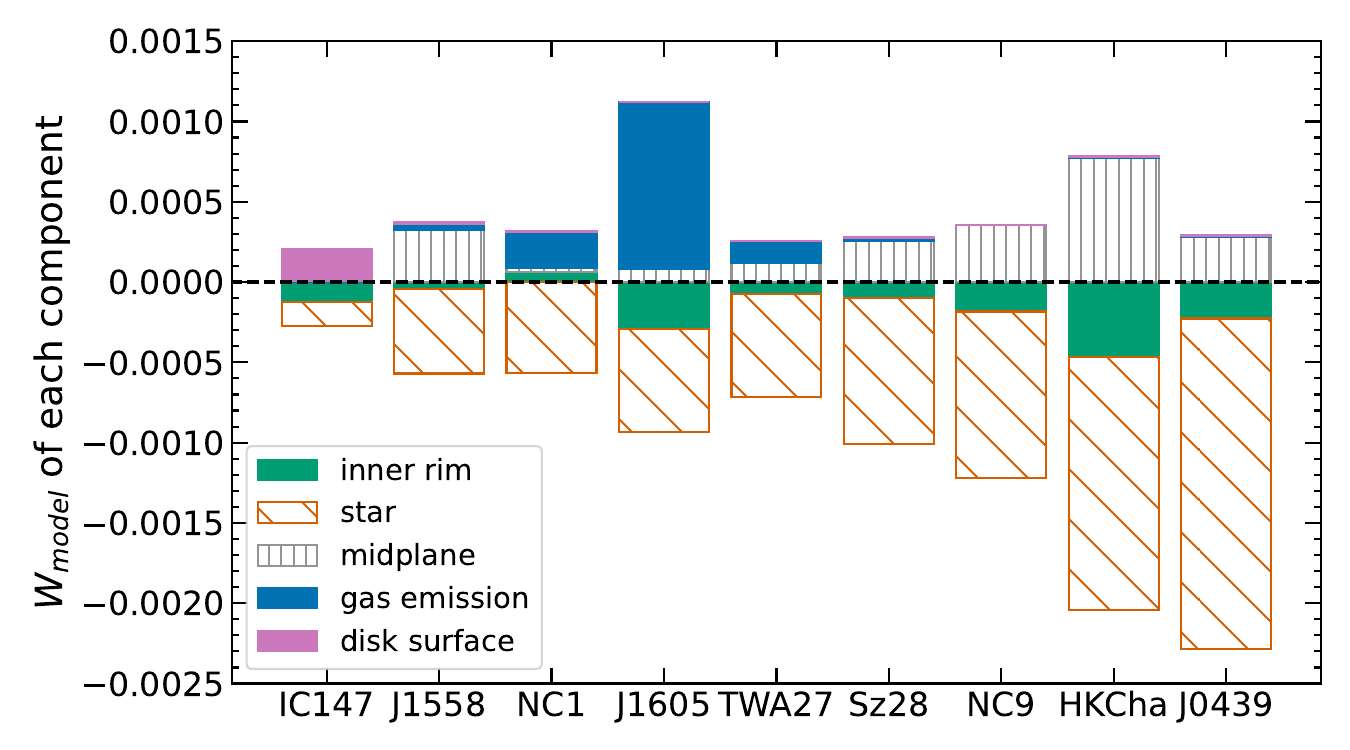}
    \caption{Contribution of each DuCKLinG model component to the depth of the water absorption feature ($W$). Negative bars represent components that produce and enhance the water absorption, while positive bars indicate components that dilute the feature. Sources are ordered from the shallowest to the deepest $W$ measured in the stellar spectra.  }
    \label{fig:water_diskcomp}
\end{figure}

\subsection{Less-, more-, and fully-settled  disks}
\paragraph{}
Our VLMS disks show a range of spectral slopes, and the slopes correlate with the $10~\mu$m silicate band strength and anti-correlate with gas column density. We also identify their dust compositions and sizes. NC1 and J1558 present the highest spectral slopes and prominent $10~\mu$m and $20~\mu$m silicate bands with weak crystalline dust features. NC9, J0439, HKCha, and Sz28 have relatively flat spectral slopes and strong crystalline features. IC147, TWA27, and J1605 show weak or absent silicate dust emission and negative spectral slopes. 

Based on these features, our sample can be categorized: (a) less-settled disks for NC1 and J1558, (b) more-settled disks for NC9, J0439, HKCha, and Sz28, and (c) fully-settled disks for IC147, TWA27, and J1605. These categories reflect overall spectral trends in the VLMS disks, and individual cases may lie between classes. For example, Sz28 and IC147 may fall between categories (b) and (c). Sz28 exhibits distinguishable silicate bands but has a negative spectral slope. IC147 has a slightly larger spectral slope than Sz28, but its weak silicate bands can only be revealed through dust retrieval modeling and extinction correction.

The less-settled disks have strong 10 and 20 $\mu$m silicate bands and therefore are expected to be dominated by small grains ($0.1-1~\mu$m) or to be abundant in $\mu$m-sized dust ($0.1-5~\mu$m). According to DuCKLinG results, they are dominated by 5 $\mu$m-sized amorphous silicates (>60 \%). Thus, the strong silicate strength in less-settled disks is due to more abundant $\mu$m-sized dust in the disk surface rather than the population of small grains. It also suggests that vertical mixing due to strong turbulence maintains large grains in the disk surface. Therefore, 5 $\mu$m grains can easily stay in the disk surface, and they effectively dominate the mass fraction compared to smaller dust grains.

The more-settled disks have weaker silicate strength than the less-settled disks despite smaller grain population, and this can be due to a lower abundance of overall $\mu$m-sized grains. Reduced disk turbulence settles large dust grains first, so the more-settled disks show lower mass fractions of large grains compared to the less-settled disks in the disk surface. Alternatively, inner-disk clearing could also explain the observed small grains. Because larger grains are more abundant in the inner disk than the outer disk, the removal of large grains in the inner disk leaves smaller grains in the outer disks. Parent-body collisions can also explain the abundant small grains. Collisions of large bodies (e.g., planetesimals) can produce sub-$\mu$m to a few $\mu$m-size dust grains that radiate warm emission above an optical depth of 1 in the disk surface \citep{Swinkels_Dominik2024}. 

To investigate the inner-disk clearing, we integrated the flux of MIRI and stellar spectra from $4.9\;\mu\rm m$ to $6.3\;\mu\rm m$, where the 10 $\mu$m silicate band and strong hydrocarbon emissions do not appear, to measure the disk-to-star flux ratios. The disk flux is obtained by subtracting the modeled stellar flux from the MIRI flux. NC1 and J1558 have ratios of 1.1 and 1.7, while NC9, J0439, HKCha, Sz28, and TWA27 have ratios of 0.9, 0.5, 1.2, 0.8, and 0.6, respectively. The inner disk emission tends to be weak as a disk is more settled. Although this is not a direct measurement of inner-disk clearing, the overall trend that more- or fully-settled disks have lower ratios than less-settled disk supports the interpretation of inner-disk clearing. From this trend, we expect all of the fully-settled disks to have even lower disk-to-star flux ratios, but the ratios for IC147 and J1605 are 2.0 and 1.4, respectively. This can be due to very large column of warm/hot hydrocarbon gas in combination with low dust opacity. For J1605, its flux is reported to be variable between the MIRI and \textit{Spitzer} data, and the MIRI data has significantly higher flux level than the \textit{Spitzer} data \citep{Tabone_etal2023}. This variability may also cause the high disk-to-star flux ratio. 

IC147, TWA27, and J1605 show very weak or absent 10 $\mu$m silicate bands and negative spectral slopes. Their dust grains are likely more evolved to larger size and settled deeper into the midplane than more-settled disks, so the mid-IR flux is dominated by gas emission and little from dust in the disk surface. As a result, the optical depth is lower, and the observed gas column density appears higher than the more-settled disks. 

\subsection{Crystalline silicates}
\paragraph{}
Crystalline mass fractions derived from DuCKLinG are shown in \fg{crystal}. NC1 and J1558 have the lowest crystallinities ($\sim 15 \%$), and NC9, J0439, HKCha, and Sz28 have higher crystallinities (> 30 \%). Sz28 has the highest crystallinity $\sim 40 \%$ (Table \ref{tab:mass_frac}). This trend is driven by enstatite, as forsterite remains relatively constant ($\sim20$ \%) across disks. We remind the reader again that these estimates apply only to small grains ($\lesssim 5\mu$m) in the disk surface detectable with MIRI. IC147, TWA27, and J1605 are excluded from this analysis due to their weak or absent silicate features, which make it difficult to study the detailed dust composition.

\begin{figure}
    \centering
    \includegraphics[width=\linewidth]{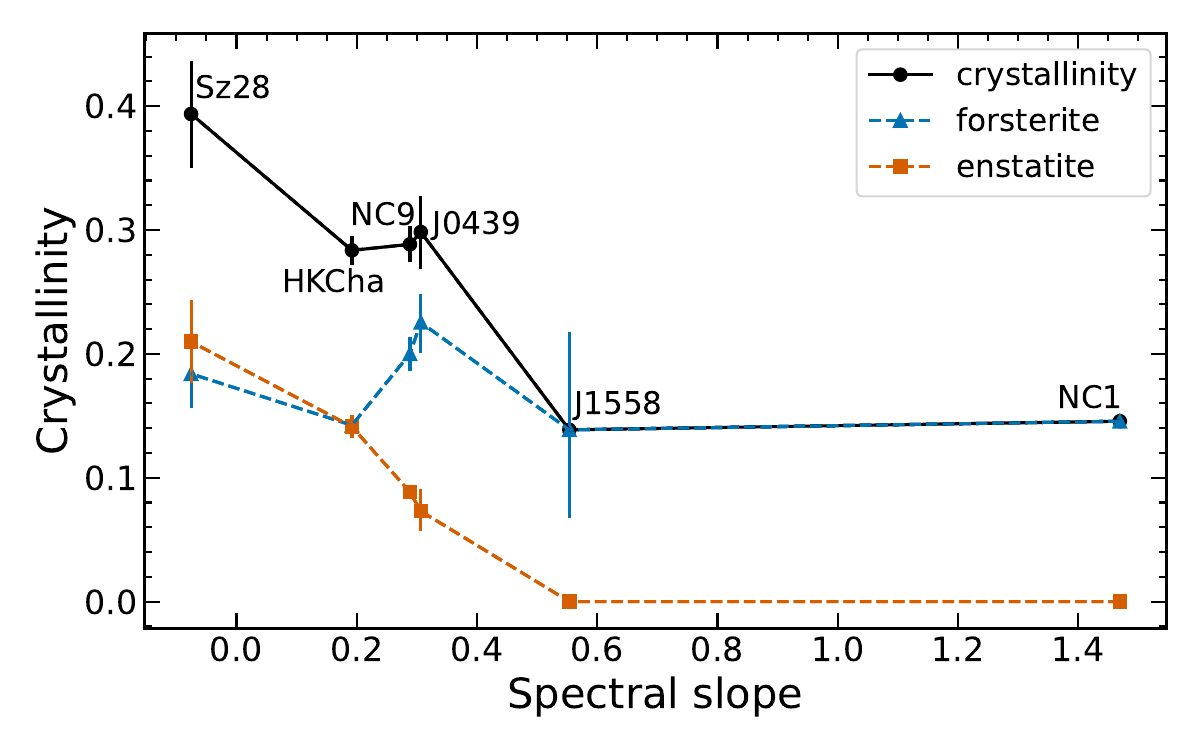}
    \caption{Crystallinity for NC1, J1558, NC9, J0439, HKCha, and Sz28. Crystallinity for IC147, TWA27, and J1605 are not shown due to their weak silicate features. The total crystallinity is shown in black, and crystallinities for forsterite and enstatite are in blue and orange dashed lines, respectively. }
    \label{fig:crystal}
\end{figure}

In Herbig disks, enstatite tends to appear at the shorter wavelengths (7-17 $\mu$m) while the longer wavelengths (17-35 $\mu$m) are more forsterite-dominated \citep{Juhasz_etal2010}. For T Tauri disks, the inner disk ($\sim$1 au; $8-13~\mu$m) is dominated by enstatite, and the outer disk ($\sim5-15$ au; $17-35.5~\mu$m) is dominated by forsterite \citep{Bouwman_etal2008}. Our sample appears to follow the same trend. Less-settled disks exhibit higher flux at longer wavelengths, which are sensitive to cooler regions in the disk and therefore trace the forsterite-dominated outer disk. In contrast, more-settled disks are dominated by shorter-wavelength emission, tracing the warmer inner disk enriched in enstatite. NC1 and J1558 that are more sensitive to the outer disk show only forsterite features, while NC9, J0439, HKCha, and Sz28 show enstatite features from the inner disk in addition to the forsterite. Therefore, there emerges a consistent picture of the distribution of crystalline silicate composition across stellar masses. 

The crystallinity increases with decreasing spectral slope. If all disks were in the same evolutionary stage, disks with higher spectral slopes should be hotter than those with lower slopes because of the abundant $\mu$m-sized dust grains in the disk surface, which efficiently absorb stellar radiation. This can cause efficient thermal annealing and hence higher crystallinity. However, contrary to this simple picture, we detect a lower crystallinity for less-settled disks (high spectral slope). This can be explained in two ways. One is that the inner disk dominates the disk emission in a settled disk due to a steeper temperature gradient. Thus, the emission from enstatite naturally increases without the change in the distribution of crystalline silicate composition discussed above. The other one is that the overall crystallinity of disks may also increase with time as radially drifting grains are annealed in the inner disk and crystalline silicates of second generation grains are produced by parent-body processing, such as collisions of planetesimals. This suggests that J0439 and NC9, in addition to NC1 and J1558, may be at an earlier stage of disk evolution than Sz 28 and HK Cha. As a result, the dust has not experienced thermal annealing as much as other more-settled disks.

\section{Conclusion}
\label{sec:conclusion}
\paragraph{}
We analyzed the dust features in the spectra of the VLMS disks observed with JWST/MIRI MRS in the MINDS program to investigate the geometrical structure, dust composition, grain sizes, and correlations with gas column density. In general, VLMS disks have weak 10 $\mu$m silicate bands. As the silicate band weakens, the spectral slope decreases. Disks with low spectral slopes tend to show higher gas column densities. Moreover, we used DuCKLinG to analyze the dust emission in the presence of a strong hydrocarbon pseudo-continuum. This allows us to decompose the spectral components and quantify the mass fractions of dust grain sizes and species. Our main conclusions are summarized as follows:

\begin{itemize}
    \item[$\bullet$] Our VLMS disks are more evolved than some T Tauri disks, and it is similar to previously found trends. While T Tauri disks from \textit{Spitzer} observations show strong 10 $\mu$m silicate bands up to F9.8 = 3, the VLMS disks have F9.8 values only up to 1.7. Moreover, F9.8/F11.3 ratios below 1.1 for the VLMS disks suggest large dust grains or highly crystalline dust, whereas T Tauri disks can exhibit higher values up to 1.6. 

    \item[$\bullet$] Less-settled disks exhibit strong silicate emission from their disk surfaces. Large F9.8 values and high spectral slopes indicate the presence of abundant $\mu$m-sized silicate grains in the disk surface.

    \item[$\bullet$] The gas column densities tend to increase as dust optical depth decreases. We find a weak correlation between the $F_{^{13}\rm CCH_2}/F_{\rm C_{2}H_{2}}$ and spectral slope, and no source with a small spectral slope shows a low gas column density. In more-settled disks, reduced dust optical depth allows observations to probe deeper into the disk, which reveals higher gas column densities. 

    \item[$\bullet$] Emission from the disk weakens the water absorption feature from the stellar spectrum. For NC1 and J1605, gas emission from HCN and C$_2$H$_2$ effectively diminishes the depth of the stellar water absorption while the midplane component generally weakens the depth of absorption. 

    \item[$\bullet$] Disk settling affects the dominant crystalline silicate species in our spectra. Less-settled disks emit forsterite more from the outer disk than more-settled disks, which allows the detection of enstatite from the inner disk.

    \item[$\bullet$] The crystallinity increases as the spectral slope decreases, and enstatite determines the increase. With decreasing slope, the crystallinity of enstatite increases while forsterite is relatively constant. 

    \item[$\bullet$] Our sample shows that the distribution of crystalline silicate is independent of stellar mass. Our sample has enstatite-rich inner disks and forsterite-rich outer disks, similar to Herbig and T Tauri disks. 
    
    \item[$\bullet$] Less-settled disks may represent earlier evolutionary stages with limited thermal annealing while more-settled disks represent more evolved systems which experienced longer thermal annealing.

\end{itemize}

In the MINDS sample, we analyzed ten VLMS disks with the exception of J0438 due to its high inclination. JWST/MIRI MRS observed more VLMS disks beyond this program, and these observations would improve the statistical significance of our findings and provide deeper insight into the co-evolution of dust and gas. 

\begin{acknowledgements}
This work is based on observations made with the NASA/ESA/CSA James Webb Space Telescope. The data were obtained from the Mikulski Archive for Space Telescopes at the Space Telescope Science Institute, which is operated by the Association of Universities for Research in Astronomy, Inc., under NASA contract NAS 5-03127 for JWST. These observations are associated with program \#1282. The following National and International Funding Agencies funded and supported the MIRI development: NASA; ESA; Belgian Science Policy Office (BELSPO); Centre Nationale d’Etudes Spatiales (CNES); Danish National Space Centre; Deutsches Zentrum fur Luft- und Raumfahrt (DLR); Enterprise Ireland; Ministerio De Econom\'ia y Competividad; Netherlands Research School for Astronomy (NOVA); Netherlands Organisation for Scientific Research (NWO); Science and Technology Facilities Council; Swiss Space Office; Swedish National Space Agency; and UK Space Agency.

We acknowledge to Combined Atlas of Sources with Spitzer IRS Spectra (CASSIS) database for the use of low-resolution spectra. 

This publication makes use of VOSA, developed under the Spanish Virtual Observatory (https://svo.cab.inta-csic.es) project funded by MCIN/AEI/10.13039/501100011033/ through grant PID2020-112949GB-I00.
VOSA has been partially updated by using funding from the European Union's Horizon 2020 Research and Innovation Programme, under Grant Agreement nº 776403 (EXOPLANETS-A)

T.K. acknowledges support from STFC Grant ST/Y002415/1.
\end{acknowledgements}

\bibliographystyle{aa}
\bibliography{ref}

\begin{appendix}
\onecolumn
\section{Stellar photospheric spectra}
\paragraph{}
We model stellar photospheric spectra of our VLMS sample with photometric data from \textit{Gaia} DR3, 2MASS, and WISE in VOSA. These photometric data are shown in black markers, and their extinction corrected data are in green markers in \fg{SED}. Resulting stellar parameters is summarized in Table \ref{tab:VOSA}.
\begin{figure}[h!]
    \centering
    \includegraphics[width=0.9\linewidth]{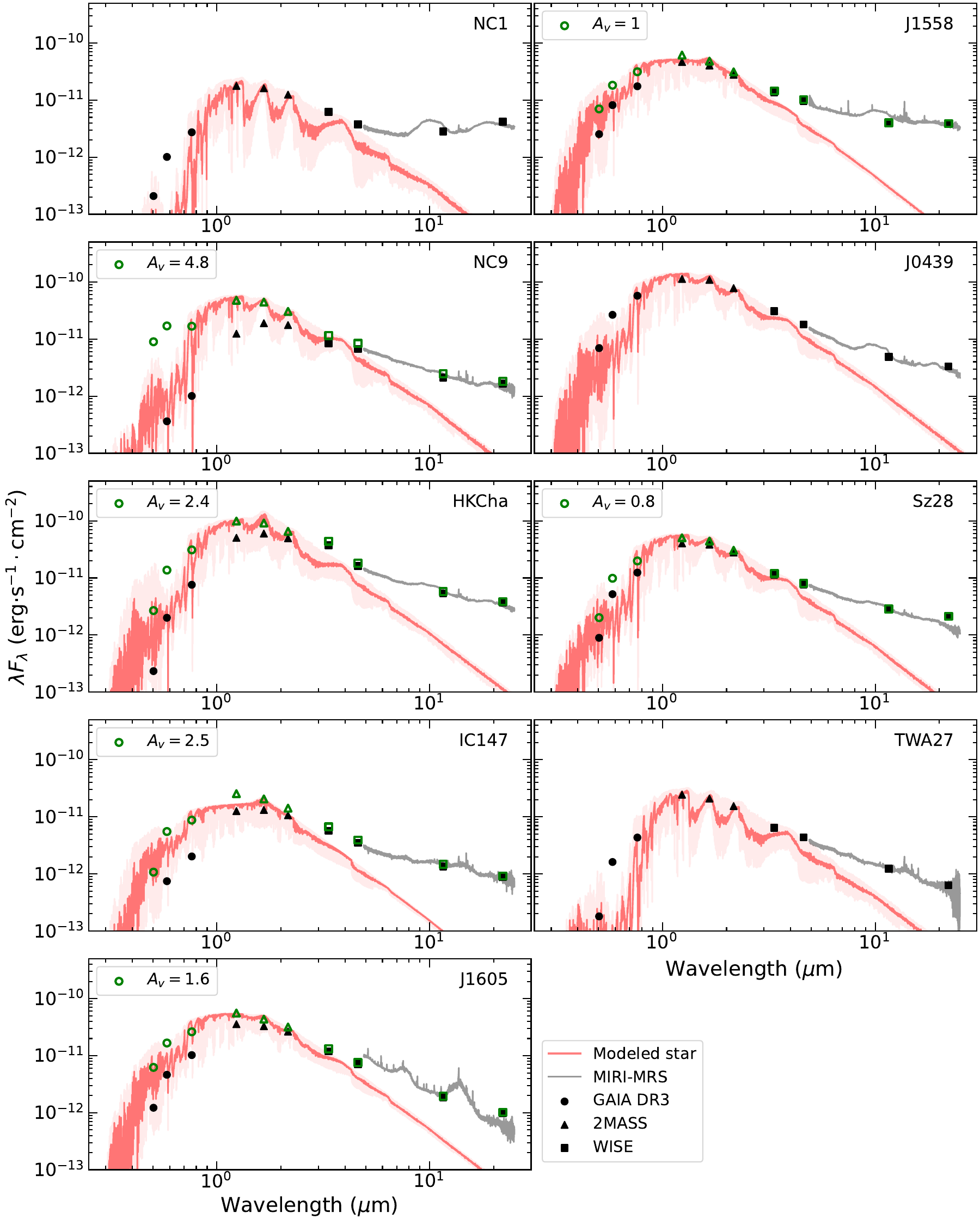}
    \caption{Modeled stellar photospheric spectra with \textit{Gaia} DR3, 2MASS, and WISE in VOSA. The observed photometric data points are the black markers while the green markers are the foreground extinction corrected ones. The shaded red lines show the full resolution stellar spectra, and the red solid lines are rebinned to $R\sim1000$. The gray line is the MIRI spectrum. }
    \label{fig:SED}
\end{figure}
\FloatBarrier

\begin{table*}[h!]
\centering
\caption{Chi-square fit results of BT-Settl model for stellar photospheric spectra in VOSA.  }
\label{tab:VOSA}
\begin{tabular}{cccccccccc}
\hline\hline
2MASS & Name & SpTy & $M_{*}/M_{\odot}$ & Distance [pc] & $A_{\rm v}$ [mag] & $T_{\rm eff}$ [K] & $\log(g)^{*}$ & $L_{\rm bol}$/$L_{\odot}$ & $\Delta$L$_{\rm bol}$/L$_{\odot}$ \\ \hline
J11071668--7735532 & NC1 & M7.74 &  $0.05^{(1)}$ & 194.6 & 0 & 2400 & 4 & 2.020e-2 & 9.704e-4 \\
J15582981--2310077 & J1558 &  M4.5 &  0.14 & 141.1  & 1 & 3600 & 4 & 4.181e-2 & 7.151e-4 \\
J11071860--7732516 & NC9 &  M5.5 &  0.08 & 197.5  & 4.8 & 2800 & 5 & 6.512e-2 & 6.261e-3 \\
J04390163+2336029 & J0439 &  M6 &  0.12 & 126.8  & 0 & 3100 & 4 & 7.212e-2 & 7.466e-4 \\
J11074245--7733593 & HKCha &  M5.25 &  0.09 & 191.0  & 2.4 & 3100 & 2.5 & 1.214e-1 & 3.271e-4 \\
J11085090--7625135 & Sz28 &  M5.25 &  0.08 & 192.2  & 0.8 & 3000 & 4.5 & 6.593e-2 & 1.281e-3 \\
J11082650--7715550 & IC147 &  M5.75 & 0.07 & 195.8  & 2.5 & 3700 & 3 & 2.886e-2 & 1.694e-3 \\
J12073346--3932539 & TWA27 &  M9 &  $0.02^{(2)}$ & 64.4  & 0 & 2500 & 3 & 3.025e-3 & 6.101e-5 \\
J16053215--1933159 & J1605 &  M4.5 &  $0.13^{(3)}$ & 152.3  & 1.6 & 3400 & 4.5 & 4.626e-2 & 9.038e-4 \\
\hline
\end{tabular}
\tablefoot{Distances are collected from \textit{Gaia} DR3 (Gaia Collaboration et al. 2021), and stellar properties, including $A_{\rm v}$, are obtained from \cite{Manara_etal2023}, which provides collected data from \cite{Manara_etal2017,Herczeg_Hillenbrand2014,Herczeg_Hillenbrand2008}. Values with superscripts are obtained from (1)\cite{Luhman2007}, (2) \cite{Manjavacas_etal2024}, and (3) \cite{Testi_etal2022}. $^{*}$The values are uncertain because they are fitted based on SED shapes instead of using a detailed spectroscopic model. }
\end{table*}

In the modeled stellar photospheric, water absorption feature appears around $6.5~\mu$m, and it corresponds to the step-like feature shown in the MIRI spectra. \Fg{J0439water} shows the water absorption feature in both stellar and MIRI spectra for J0439 as an example. 

\begin{figure}[h]
    \centering
    \includegraphics[width=0.9\linewidth]{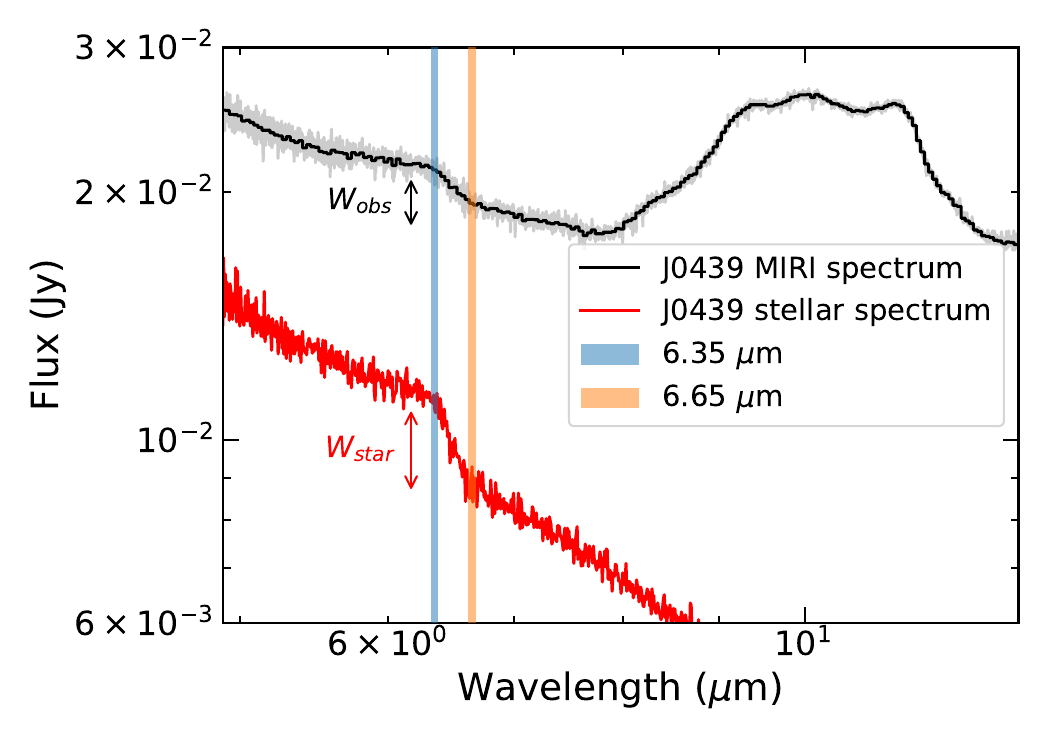}
    \caption{Example of water absorption feature in the modeled photospheric spectrum and the MIRI spectrum of J0439. The step-like feature around 6.5 $\mu$m aligns with the water absorption feature in the modeled photospheric spectrum. The depth of the absorption is measured using the fluxes at 6.35 $\mu$m (blue shaded region) and 6.65 $\mu$m (orange shaded region).}
    \label{fig:J0439water}
\end{figure}

\section{DuCKLinG models for different spectral resolution}
\label{sec:R200}
\paragraph{}
In this section, we compare DuCKLinG results for different spectral resolutions of the Sz 28 spectrum that is already well studied in \cite{Kaeufer_etal2024}: (a) fully rebinned to $R\sim 200$ and (b) generally rebinned to $R\sim 200$ with full MIRI resolution from $11.7~\mu$m to $16.8~\mu$m. (b) is the one what we decided to use in this study. 

In the case of (a), DuCKLinG finds CO$_2$ to be present from $9$ to $19~\mu$m as shown in \fg{Sz28gas} while we know that the CO$_2$ emission only contributes in the $14~\mu$m region in (b). The column density and emitting temperature of CO$_2$ are $10^{23}$ cm$^{-2}$ and $619^{+390}_{-134}$ K in (a) while they are $10^{19}$ cm$^{-2}$ and $311.7^{+138}_{-89}$~K in (b). \cite{Kaeufer_etal2024} also used DuCKLinG to fit gas emission using the full resolution of MIRI, and they report $9\times10^{18}$ cm$^{-2}$ and $203-241$ K. This shows that (a) significantly overestimates the column density and temperature of CO$_2$. This produces substructures in the dust continuum, and it results in DuCKLinG detecting small crystalline silicates as shown in \fg{Sz28a}. In (a), the gas features are smeared out so that the fit optimizes for the residual not at the peaks but over large wavelength ranges. This results in many molecules to be wrongly treated as quasi-continua. The estimates from approach (b) are not exactly the same as in \cite{Kaeufer_etal2024} but much closer. The remaining difference comes from the simpler fitting of the gas emissions in this study. Therefore, using the high resolution in the $14~\mu$m region is important for reasonable fitting of the gas emission which is essential not to misinterpret dust features.    
\begin{figure}[h!]
    \centering
    \includegraphics[width=0.8\linewidth]{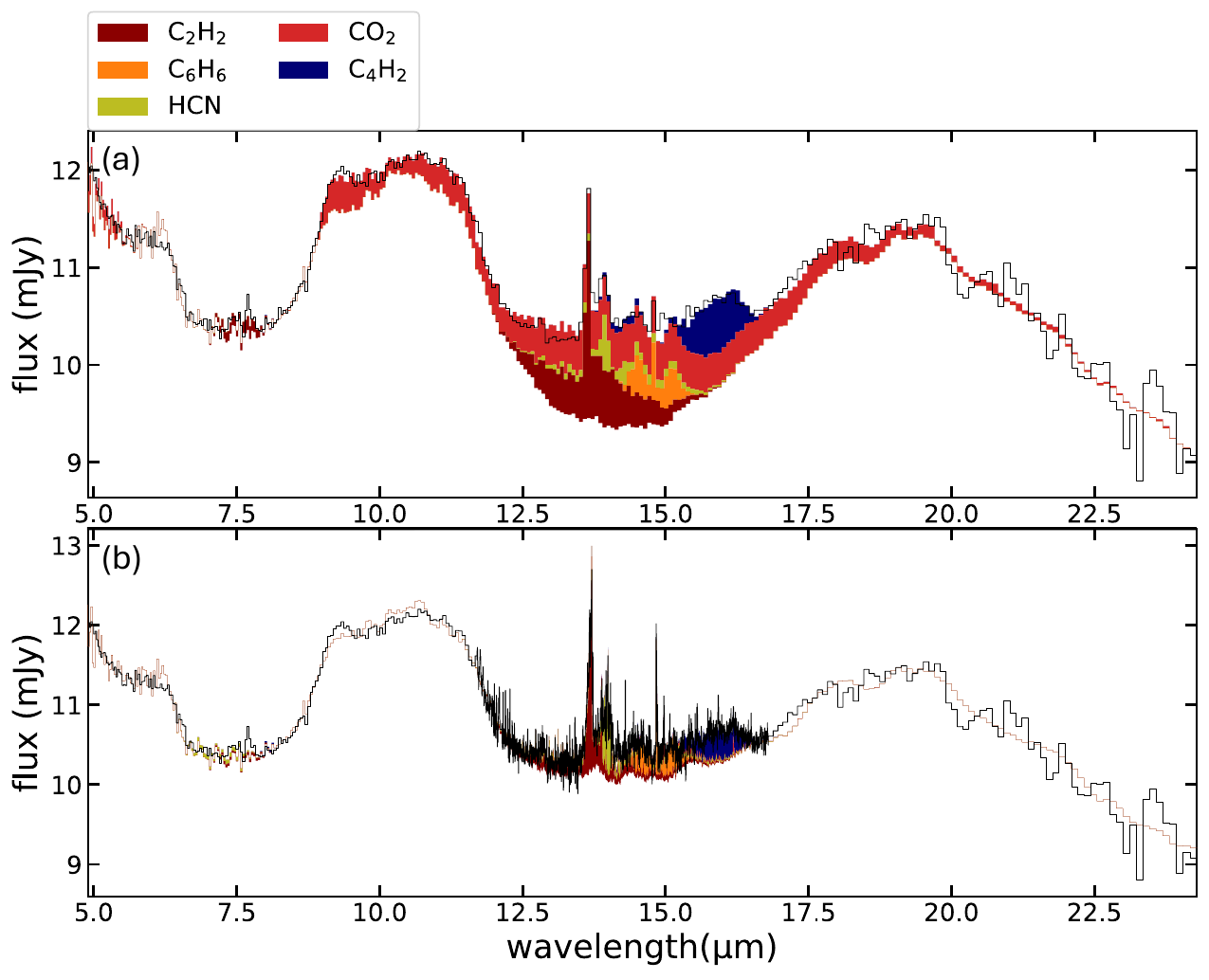}
    \caption{Gas emission contributions for Sz28 spectrum. (a) Sz28 spectrum is fully rebinned to $R\sim 200$. (b) The spectrum is also rebinned to $R\sim 200$, but the $14~\mu$m region is full MIRI resolution. }
    \label{fig:Sz28gas}
\end{figure}

\begin{figure}[]
    \centering
    \includegraphics[width=0.8\textwidth]{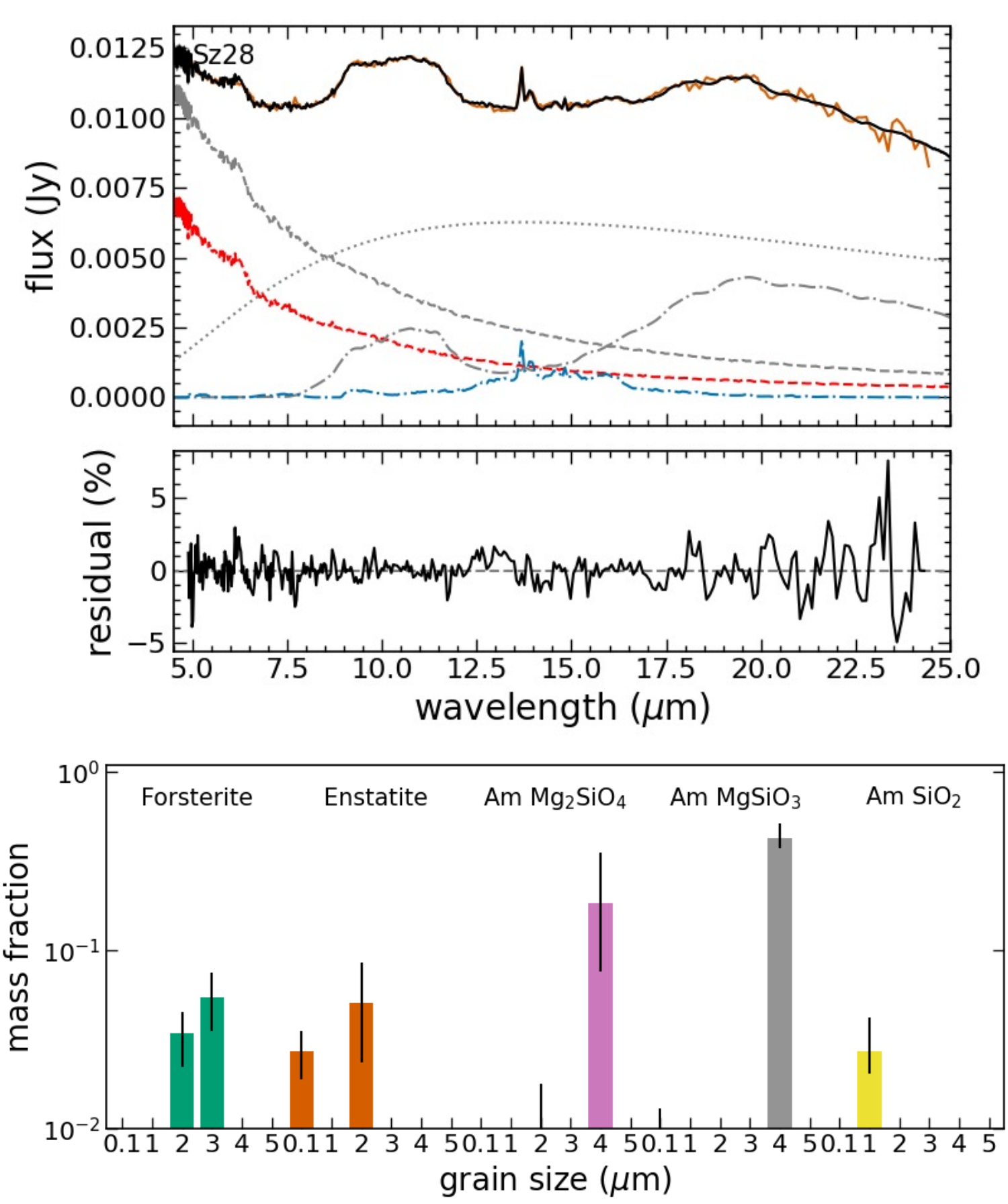}
    \caption{DuCKLinG result for Sz28 of $R\sim 200$. Mass fractions of each dust species and grain sizes are indicated with their errorbars.}
    \label{fig:Sz28a}
\end{figure}
\FloatBarrier

\section{Mass fractions of DuCKLinG models}
\label{sec:masstable}
\paragraph{}
The results of DuCKLinG models for our VLMS disks, mass fractions for dust species and grain sizes for each disk, are summarized in Table \ref{tab:mass_frac} and Table \ref{tab:mass_frac2}. Table \ref{tab:mass_frac} is for models with forsterite, enstatite, amorphous Mg$_2$SiO$_4$, MgSiO$_3$, and SiO$_2$, and gas species include C$_2$H$_2$, C$_6$H$_6$, HCN, CO$_2$, and C$_4$H$_2$. Table \ref{tab:mass_frac2} is for models that additionally use an arbitrary flat dust opacity and C$_2$H$_4$ gas emission. 

\begin{table*}[]
\centering
\caption{Mass fractions of dust species and grain sizes for each disk. }
\label{tab:mass_frac}
\begin{tabular}{ccccccc}
\hline\hline
grain size & 0.1 $\mu$m & 1 $\mu$m & 2 $\mu$m & 3 $\mu$m & 4 $\mu$m & 5 $\mu$m \\ \hline
&&&NC1 (\%)&&& \\ \hline
Am Mg$_2$SiO$_4$ & 0 & 0 & 0 & 0 & $0.00^{+2.52}_{-0.00}$ & $26.17^{+0.55}_{-2.74}$ \\
Am MgSiO$_3$ & $2.90^{+0.68}_{-0.61}$ & $9.10^{+0.64}_{-0.83}$ & $0.00^{+0.22}_{-0.00}$ & 0 & $0.00^{+2.79}_{-0.00}$ & $46.66^{+0.82}_{-2.64}$ \\
Am SiO$_2$ & 0 & 0 & 0 & 0 & 0 & 0 \\
Forsterite & 0 & $0.54^{+0.04}_{-0.04}$ & 0 & 0 & 0 & $14.02^{+0.36}_{-0.36}$ \\
Enstatite & 0 & 0 & 0 & 0 & 0 & 0 \\ \hline
&&&J1558 (\%)&&& \\ \hline
Am Mg$_2$SiO$_4$ & 0 & 0 & 0 & 0 & 0 & $55.69^{+1.11}_{-1.96}$ \\
Am MgSiO$_3$ & 0 & $4.28^{+1.30}_{-2.75}$ & $0.00^{+9.46}_{-0.00}$ & $18.74^{+3.07}_{-10.43}$ & 0 & $2.21^{+7.68}_{-2.21}$ \\
Am SiO$_2$ & 0 & 0 & 0 & 0 & 0 & $2.95^{+0.51}_{-0.48}$ \\
Forsterite & 0 & $1.71^{+0.74}_{-0.51}$ & 0 & $5.80^{+3.65}_{-5.80}$ & 0 & $6.35^{+6.98}_{-4.05}$ \\
Enstatite & 0 & 0 & 0 & 0 & 0 & 0 \\ \hline
&&&NC9 (\%)&&& \\ \hline
Am Mg$_2$SiO$_4$ & $25.96^{+1.54}_{-1.82}$ & 0 & 0 & 0 & 0 & $0.00^{+0.55}_{-0.00}$ \\
Am MgSiO$_3$ & 0 & $36.06^{+2.32}_{-2.84}$ & 0 & 0 & 0 & $5.72^{+5.14}_{-4.57}$ \\
Am SiO$_2$ & $2.30^{+0.40}_{-0.58}$ & $0.52^{+0.72}_{-0.52}$ & 0 & 0 & 0 & 0 \\
Forsterite & $1.80^{+0.41}_{-0.38}$ & 0 & $4.94^{+1.22}_{-1.20}$ & $13.27^{+1.92}_{-2.11}$ & 0 & 0 \\
Enstatite & 0 & $6.76^{+0.46}_{-0.54}$ & 0 & 0 & 0 & $2.08^{+1.77}_{-1.55}$ \\ \hline
&&&J0439 (\%)&&& \\ \hline
Am Mg$_2$SiO$_4$ & $15.02^{+1.50}_{-1.99}$ & 0 & $0.00^{+2.65}_{-0.00}$ & $0.00^{+13.44}_{-0.00}$ & $12.23^{+3.07}_{-12.23}$ & 0 \\
Am MgSiO$_3$ & $12.85^{+0.58}_{-0.52}$ & 0 & 0 & 0 & $25.14^{+0.82}_{-1.05}$ & 0 \\
Am SiO$_2$ & 0 & $1.67^{+0.11}_{-0.12}$ & 0 & 0 & 0 & 0 \\
Forsterite & 0 & $8.86^{+0.17}_{-0.18}$ & $5.23^{+0.85}_{-0.90}$ & $8.45^{+0.99}_{-0.98}$ & 0 & $0.00^{+0.35}_{-0.00}$ \\
Enstatite & $1.40^{+0.17}_{-0.20}$ & $0.00^{+0.12}_{-0.00}$ & $5.90^{+0.43}_{-0.53}$ & 0 & 0 & 0 \\ \hline
&&&HKCha (\%)&&& \\ \hline
Am Mg$_2$SiO$_4$ & $9.24^{+1.62}_{-1.26}$ & 0 & 0 & 0 & $0.00^{+3.52}_{-0.00}$ & $23.70^{+2.23}_{-3.04}$ \\
Am MgSiO$_3$ & 0 & $18.89^{+0.91}_{-1.05}$ & $9.75^{+1.95}_{-1.82}$ & 0 & $0.00^{+2.23}_{-0.00}$ & $5.97^{+2.22}_{-4.23}$ \\
Am SiO$_2$ & 0 & $2.54^{+0.13}_{-0.12}$ & 0 & 0 & 0 & 0 \\
Forsterite & 0 & 0 & $6.49^{+0.29}_{-0.45}$ & $7.75^{+0.68}_{-0.67}$ & 0 & 0 \\
Enstatite & 0 & 0 & $4.55^{+0.29}_{-0.29}$ & 0 & 0 & $9.57^{+0.88}_{-0.86}$ \\ \hline
&&&Sz28 (\%)&&& \\ \hline
Am Mg$_2$SiO$_4$ & $0.00^{+0.96}_{-0.00}$ & 0 & $17.27^{+1.86}_{-8.99}$ & 0 & $0.00^{+10.39}_{-0.00}$ & 0 \\
Am MgSiO$_3$ & $3.21^{+0.91}_{-0.92}$ & 0 & 0 & 0 & $34.84^{+1.82}_{-2.10}$ & 0 \\
Am SiO$_2$ & 0 & $3.00^{+0.18}_{-0.31}$ & 0 & 0 & 0 & 0 \\
Forsterite & 0 & $0.34^{+0.46}_{-0.34}$ & $0.47^{+1.19}_{-0.47}$ & 0 & $17.59^{+2.31}_{-2.68}$ & 0 \\
Enstatite & 0 & 0 & $13.09^{+2.18}_{-1.83}$ & 0 & $7.89^{+2.59}_{-2.81}$ & 0 \\ \hline
&&&IC147 (\%)&&& \\ \hline
Am Mg$_2$SiO$_4$ & 0 & $25.22^{+2.99}_{-5.10}$ & $0.00^{+7.92}_{-0.00}$ & 0 & 0 & 0 \\
Am MgSiO$_3$ & 0 & 0 & 0 & 0 & 0 & $10.03^{+2.53}_{-2.64}$ \\
Am SiO$_2$ & $0.12^{+0.12}_{-0.12}$ & 0 & 0 & 0 & 0 & 0 \\
Forsterite & 0 & 0 & 0 & 0 & 0 & $57.75^{+1.97}_{-1.85}$ \\
Enstatite & 0 & 0 & 0 & 0 & 0 & $5.06^{+2.73}_{-2.87}$ \\ \hline
&&&TWA27 (\%)&&& \\ \hline
Am Mg$_2$SiO$_4$ & 0 & 0 & 0 & 0 & 0 & 0 \\
Am MgSiO$_3$ & 0 & 0 & 0 & 0 & 0 & 0 \\
Am SiO$_2$ & $14.37^{+1.38}_{-1.19}$ & 0 & 0 & 0 & 0 & 0 \\
Forsterite & 0 & 0 & 0 & 0 & 0 & 0 \\
Enstatite & 0 & 0 & 0 & 0 & $79.78^{+5.62}_{-20.32}$ & $4.40^{+19.06}_{-4.40}$ \\ \hline
&&&J1605 (\%)&&& \\ \hline
Am Mg$_2$SiO$_4$ & 0 & 0 & 0 & 0 & 0 & 0 \\
Am MgSiO$_3$ & 0 & 0 & 0 & 0 & 0 & 0 \\
Am SiO$_2$ & $0.06^{+0.24}_{-0.06}$ & 0 & 0 & 0 & 0 & 0 \\
Forsterite & 0 & 0 & 0 & 0 & $0.76^{+5.93}_{-0.76}$ & $99.08^{+0.92}_{-5.90}$ \\
Enstatite & 0 & 0 & 0 & 0 & 0 & 0 \\
\hline
\end{tabular}
\end{table*}

\begin{table*}[]
\centering
\caption{Mass fractions of dust species and grain sizes for IC147, TWA27, J1605 of DuCKLinG model with the flat opacity and C$_2$H$_4$ emission.}
\label{tab:mass_frac2}
\begin{tabular}{ccccccc}
\hline\hline
grain size & 0.1 $\mu$m & 1 $\mu$m & 2 $\mu$m & 3 $\mu$m & 4 $\mu$m & 5 $\mu$m \\ \hline
&&&IC147 (\%)&&& \\ \hline
Am Mg$_2$SiO$_4$ & $5.48^{+0.08}_{-0.07}$ & 0 & 0 & 0 & 0 & 0 \\
Am MgSiO$_3$ & 0 & 0 & 0 & 0 & 0 & 0 \\
Am SiO$_2$ & $0.04^{+0.01}_{-0.01}$ & 0 & 0 & 0 & 0 & 0 \\
Forsterite & 0 & 0 & 0 & $0.35^{+0.12}_{-0.14}$ & 0 & $1.14^{+0.29}_{-0.27}$ \\
Enstatite & $0.12^{+0.01}_{-0.01}$ & 0 & 0 & 0 & 0 & 0 \\
Flat opacity & - & - & - & - & - & $92.88^{+0.19}_{-0.22}$ \\\hline
&&&TWA27 (\%)&&& \\ \hline
Am Mg$_2$SiO$_4$ & 0 & 0 & 0 & 0 & 0 & 0 \\
Am MgSiO$_3$ & 0 & 0 & 0 & 0 & 0 & 0 \\
Am SiO$_2$ & $15.24^{+1.61}_{-1.41}$ & 0 & 0 & 0 & 0 & 0 \\
Forsterite & 0 & 0 & 0 & $0.00^{+5.29}_{-0.00}$ & 0 & 0 \\
Enstatite & 0 & 0 & 0 & 0 & $79.27^{+5.43}_{-7.62}$ & $0.51^{+11.27}_{-0.51}$ \\
Flat opacity & - & - & - & - & - & 0 \\\hline
&&&J1605 (\%)&&& \\ \hline
Am Mg$_2$SiO$_4$ & 0 & 0 & 0 & 0 & 0 & 0 \\
Am MgSiO$_3$ & 0 & 0 & 0 & 0 & 0 & 0 \\
Am SiO$_2$ & $98.28^{+1.72}_{-3.18}$ & 0 & 0 & 0 & 0 & 0 \\
Forsterite & $1.72^{+3.16}_{-1.72}$ & 0 & 0 & 0 & 0 & 0 \\
Enstatite & 0 & 0 & 0 & 0 & 0 & 0 \\
Flat opacity & - & - & - & - & - &  0 \\
\hline
\end{tabular}
\end{table*}

\end{appendix}
\end{document}